\documentclass[12pt,a4paper]{article}

\usepackage{lineno}  

\usepackage{xspace}

\usepackage{picture}
\usepackage{graphics}
\usepackage{graphicx}
\usepackage{color}
\usepackage{ifpdf}
\usepackage{subfigure}

\usepackage{amsmath}

\usepackage{ifthen} 

\newboolean{uprightparticles}
\setboolean{uprightparticles}{false} 
\usepackage{amssymb}
\usepackage{amsfonts}
\usepackage{upgreek}
\begin{document}



\begin{titlepage}

\vspace*{-1.5cm}

\hspace*{-0.5cm}

\vspace*{4.0cm}

{\bf\boldmath\huge
\begin{center}
Charged Particle Tracking with the Timepix ASIC
\end{center}
}

\vspace*{2.0cm}
\begin{center}
Kazuyoshi Akiba$^4$,
Marina Artuso$^{10}$,
Ryan Badman$^{10}$,
Alessandra Borgia$^{10}$,
Richard Bates$^3$,
Florian Bayer$^8$,
Martin van Beuzekom$^4$,
Jan Buytaert$^1$, 
Enric Cabruja$^2$,
Michael Campbell$^1$,
Paula Collins$^{1\dagger}$,
Michael Crossley$^{1,5}$,
Raphael Dumps$^1$,
Lars Eklund$^3$,
Daniel Esperante$^7$,
Celeste Fleta$^2$,
Abraham Gallas$^7$,
Miriam Gandelman$^6$,
Justin Garofoli$^{10}$,
Marco Gersabeck$^1$,
Vladimir V. Gligorov$^{1\dagger}$,
Hamish Gordon$^5$,
Erik H. M. Heijne$^{1,4}$,
Veerle Heijne$^4$,
Daniel Hynds$^3$,
Malcolm John$^5$,
Alexander Leflat$^{1,11}$
Lourdes Ferre Llin$^3$,
Xavi Llopart$^1$,
Manuel Lozano$^2$,
Dima Maneuski$^3$,
Thilo Michel$^8$,
Michelle Nicol$^{1,3}$,
Matt Needham$^9$,
Chris Parkes$^3$,
Giulio Pellegrini$^2$,
Richard Plackett$^{1,3}$,
Tuomas Poikela$^1$,
Eduardo Rodrigues$^3$,
Graeme Stewart$^3$,
Jianchun Wang$^{10}$,
Zhou Xing$^{10}$
\bigskip\\
{\it\footnotesize
$ ^1$Organisation Europ\'{e}enne pour la Recherche Nucl\'{e}aire, Geneva, Switzerland\\
$ ^2$Instituto de Microelectr\'{o}nica de Barcelona, IMB-CNM-CSIC, Barcelona, Spain\\
$ ^3$University of Glasgow, United Kingdom\\
$ ^4$Nationaal Instituut Voor Subatomaire Fysica, Amsterdam, Netherlands\\
$ ^5$University of Oxford, United Kingdom\\
$ ^6$Universidade Federal do Rio de Janeiro, Brazil\\
$ ^7$University of Santiago, Spain\\
$ ^8$Erlangen Centre of Astroparticle Physics, University of
Erlangen-Nuremberg, Germany\\
$ ^9$School of Physics and Astronomy, University of Edinburgh, Edinburgh, United Kingdom\\
$^{10}$Syracuse University, Syracuse, NY 13244, USA\\
$^{11}$Institute of Nuclear Physics, Moscow State University (SINP MSU), Moscow, Russia\\
}

$^\dagger$ Corresponding authors.
\end{center}

\clearpage
\begin{abstract}
  \noindent
A prototype particle tracking telescope has been constructed using 
Timepix and Medipix ASIC hybrid pixel assemblies as the
six sensing planes.  Each telescope plane consisted of one $1.4~{\rm
  cm^2}$ assembly, providing a $256 \times 256$ array of $55~{\rm
  \mu m}$ square pixels.  The telescope achieved a pointing resolution
of $2.3~{\rm \mu m}$ at the position of the device under test.  During
a beam test in 2009 the telescope
was used to evaluate in detail the performance of two Timepix hybrid pixel
assemblies; a standard planar $300~{\rm \mu m}$
thick sensor, and $285~{\rm \mu m}$ thick double sided 3D sensor. 
This paper describes a detailed charge calibration study of the pixel
devices, which allows the true charge to be extracted,
and reports on measurements
of the charge collection characteristics and Landau distributions. The
planar sensor achieved a best resolution of $4.0 \pm 0.1$~$\mu$m for
angled tracks, and resolutions of between $4.4$ and $11~\mu$m for
perpendicular tracks, depending on the applied bias voltage. 
The double sided 3D sensor, which has significantly less charge
sharing, was found to have an optimal resolution of $9.0 \pm
0.1$~$\mu$m for angled tracks, and a resolution of $16.0 \pm
0.2$~$\mu$m for perpendicular tracks.
Based on these studies it is concluded that the Timepix ASIC shows
an excellent performance when used as a device for charged particle tracking.
\end{abstract}

\vspace*{2.0cm}
\vspace{\fill}

\end{titlepage}

\pagestyle{empty}  


\newpage
\setcounter{page}{2}
\mbox{~}


\tableofcontents
\cleardoublepage


\pagestyle{plain} 
\setcounter{page}{1}
\pagenumbering{arabic}


%

\section{Introduction}
\label{s:introduction}
Pixel detectors are an attractive choice for the inner tracking
regions of current and future particle physics detectors, as they provide
high granularity, radiation hardness, and ease of pattern recognition.   A
candidate pixel ASIC which is well suited to applications at LHC,
SLHC, and future forward geometry trackers with high rate and
resolution requirements is the Timepix chip characterised in this paper.
The 55~${\rm \mu m}$ square pixel allows single sided
modules to be built, and it can supply analogue time
over threshold information in addition to the ability to associate
hits to the correct bunch crossing via time stamping.    
The Timepix chip has not been extensively investigated for the
purposes of charged particle tracking, and for this reason a test
experiment was performed in a charged 120~GeV pion beam at CERN's 
North Area.  A particle telescope was constructed using
an array of Timepix and Medipix2 silicon sensor assemblies, and this was used to
measure the performance  of a dedicated test assembly whilst varying
parameters such as silicon bias, track incident angle, and chip settings.
The telescope was also used to measure the performance of
a 3D sensor bonded to a Timepix chip, which represents an option for radiation
hard SLHC applications.  The same 3D sensor was also tested in an
X-ray beam, and the results are reported in the companion
paper~\cite{3Dcompanion}.
Due to the excellent performance of
the Timepix telescope, it was possible to perform a detailed
investigation of the resolution, efficiency and charge sharing
performance of the devices under test.  This paper reports for the
first time a detailed measurement of the resolution of a 55~$\mu$m
square pixel device
as a function of the angle of incident tracks to both the normal and
perpendicular directions with respect to the pixel columns.
The charge sharing characteristics and Landau distributions
obtained are discussed in detail.

Of particular interest are the plans to upgrade the LHCb silicon
vertex tracker (VELO~\cite{mvb})
for which extensive R\&D is being performed to develop a design based on pixel technology~\cite{LoIEoi,janvertex,hiroshima}. A
pixel readout chip dedicated to the VELO application (VeloPix) will be
developed based on Timepix, which will retain the advantageous
geometry and analogue ToT information, but have a significantly
increased readout bandwidth to cope with the higher data rates anticipated.
For this reason two sections are included in this paper
devoted to specific measurements relevant for LHC upgrades: the time
stamping capabilities of the Timepix ASIC and the effect of reducing
the number of bits available for the analogue charge measurements.

This paper is organised as follows.  The Timepix chip is introduced in Section~\ref{s:timepixchip},
followed by a description of the telescope concept and layout in Section~\ref{s:telescope}.  The
calibration of the chip, which introduces significant offsets compared
the amount of charge deposited by minimum ionising particles, is discussed in
detail.  The telescope setup is described, and the extraction of the
performance, in particular the resolution at the device under test, is
given in Sections~\ref{s:datatreatment}~and~\ref{s:analysisofdeviceundertest}.
Finally, results are presented for the devices under test:
the planar sensor, which is based on well known technology providing
information on the performance of the Timepix chip, and the double-sided 3D sensor, which
is a novel device being tested together with the Timepix chip for the first
time in a pion beam, in Sections~\ref{s:clusterchars}-\ref{s:gainsection}. The results are summarised in the
Section~\ref{s:conclusions}. 

This paper reports on work carried out jointly between the Medipix2 and LHCb
collaborations.
\section{The Timepix chip}
\label{s:timepixchip}
\subsection{Description of the chip}
The design of the Timepix~\cite{timepix} readout chip is derived from that of the Medipix2~\cite{medipix2} chip which was developed for
single photon counting applications. The development of the Timepix chip took place within the context of the Medipix2 collaboration
at the request of and with support from the EUDet Consortium~\cite{eudet}.

The Timepix ASIC comprises a $256 \times 256$ matrix of 55~$\mu$m square pixels, each of which contains its own analogue and
digital circuitry. A globally applied shutter signal determines when all pixels are active, switching between recording
data and transferring it off the chip. 
Each pixel contains a preamplifier, a discriminator with a globally adjustable
threshold, followed by mode control logic and a 14-bit pseudorandom counter with overflow logic which stops after 11,810 counts.
The threshold applied to each pixel can be individually tuned by a four-bit in-pixel trimming circuit to compensate for variations in fabrication.
The three modes
of operation are counting, time of arrival (ToA), and time over threshold (ToT).  In counting mode the Timepix pixels
behave in a similar manner to the Medipix2 pixels, incrementing the counter each time the output of the amplifier
passes the threshold.  In ToA mode the pixel records the time it was first hit. The counter is started when the amplifier
first passes the threshold and is stopped when the shutter closes. 
In this mode the depth of the counter and the overflow logic limit the shutter opening, a particular limitation at high clock frequencies. 
Beyond 11,810 counts the counter saturates. 
In ToT mode the discriminator output is used to gate the clock to the counter,
thus providing an indication of the total energy deposited. 
The clock and counter are used to record
the time the amplifier signal is above threshold. 
The amplifier has been designed so that this value is linear up to 50,000 electrons and its rise and fall time can be tuned respectively by adjusting the preamplifier and I$_\textbf{krum}$ DAC values. However, in ToT mode the dynamic range of the measurement extends well beyond the linear range of the preamplifier output.
The nominal rise time is ~100 ns and the fall time ~2 μs.
It is
possible to individually program the pixels to operate in different modes from their neighbours and be read out
simultaneously.  
The entire chip is read out after the shutter signal goes low. Using the serial readout at 200 MHz the time to read the whole matrix is $\sim$5~ms.  There is a possibility of reading out the chip using a 32-bit parallel bus. In this case the maximum clock frequency is 100 MHz and the minimum readout time $\sim$ 300~${\rm \mu s}$.
Timepix has a noise per pixel level of $\sim${100} electrons rms and a
threshold variation after tuning of $\sim${35} electrons rms, which leads to a minimum detectable signal on all the pixels of $\sim${650} electrons.

The Timepix and Medipix2 ASICs used for the telescope were bump bonded to 300~$\mu$m thick planar silicon sensor
chips which were nominally biased to 100~V.  Although Timepix was designed primarily to
operate without a sensor it retains the Medipix2 threshold trimming and sensor polarity circuitry, making it
equally well suited to reading out silicon sensors.
\subsection{Use of the chip}
In the measurements described in this paper the Timepix chips were predominantly used in ToT mode as it allowed
precise track position calculations to be performed when charge was deposited in multiple pixels by the same particle.
This gave the opportunity to reconstruct the impact position of the particle with a precision better than the
binary resolution, which would correspond to the 55~$\mu$m pitch divided by $\sqrt{12}$.

As the Timepix and Medipix2 chips are read out using a shutter signal in a ``camera'' like manner and there is no
delay line or pipeline implemented in the pixels, it is not possible to use a trigger from a scintillator to
read out a selected event. However, provided the hits are sparsely distributed across the chip several hundred
tracks can be accumulated in one frame whilst the sensor is continuously sensitive and then read out when the shutter is closed.

Before the telescope assemblies can be operated it is necessary to set the threshold trim settings for each
pixel. These bits are set to values that allow the pixels across the matrix to respond to the global threshold
in a uniform way, compensating for any variation present due to inhomogenieties in fabrication or minor
radiation damage. The process of setting these bits is referred to as the ``threshold equalisation''.

Two methods of threshold equalisation are of relevance to this report, the ``noise edge'' method, which has been
used as the standard method by the Medipix group, and the `noise centre' method, which provides a good estimation
of the global noise floor of the chip. 
Both methods operate by scanning the threshold level and binning the pixels according to the point at which they are noisy. Threshold scans are made with the trims set to their minimum and maximum values.
The optimum value of the trim
for each pixel is then determined and the threshold re-scanned to confirm the width of the distribution.  The noise
edge and noise centre methods differ in their approach by how they locate the value of the pixels' noise point. 
Using the noise mean method the threshold is scanned all the way through the noise floor and the pixels are binned according to the threshold DAC value at which the noise counts are highest. The noise edge method scans the threshold until a pixel passes a preset number of noise hits per unit time, and then uses this DAC value to bin the pixel and ceases to scan that pixel. This means that there is an offset between the absolute values of the noise position recorded. The noise mean method returns a central value which is referred to in this paper as the ``noise floor''. Typically the noise edge method returns a value with a negative offset
of about 20 THL DAC steps (the threshold decreases with increasing threshold DAC value), corresponding to roughly 500 electrons, with respect to the noise floor.
\section{The Timepix telescope}
\label{s:telescope}
\begin{figure}[h]
  \begin{center}
   \includegraphics[width=14cm]{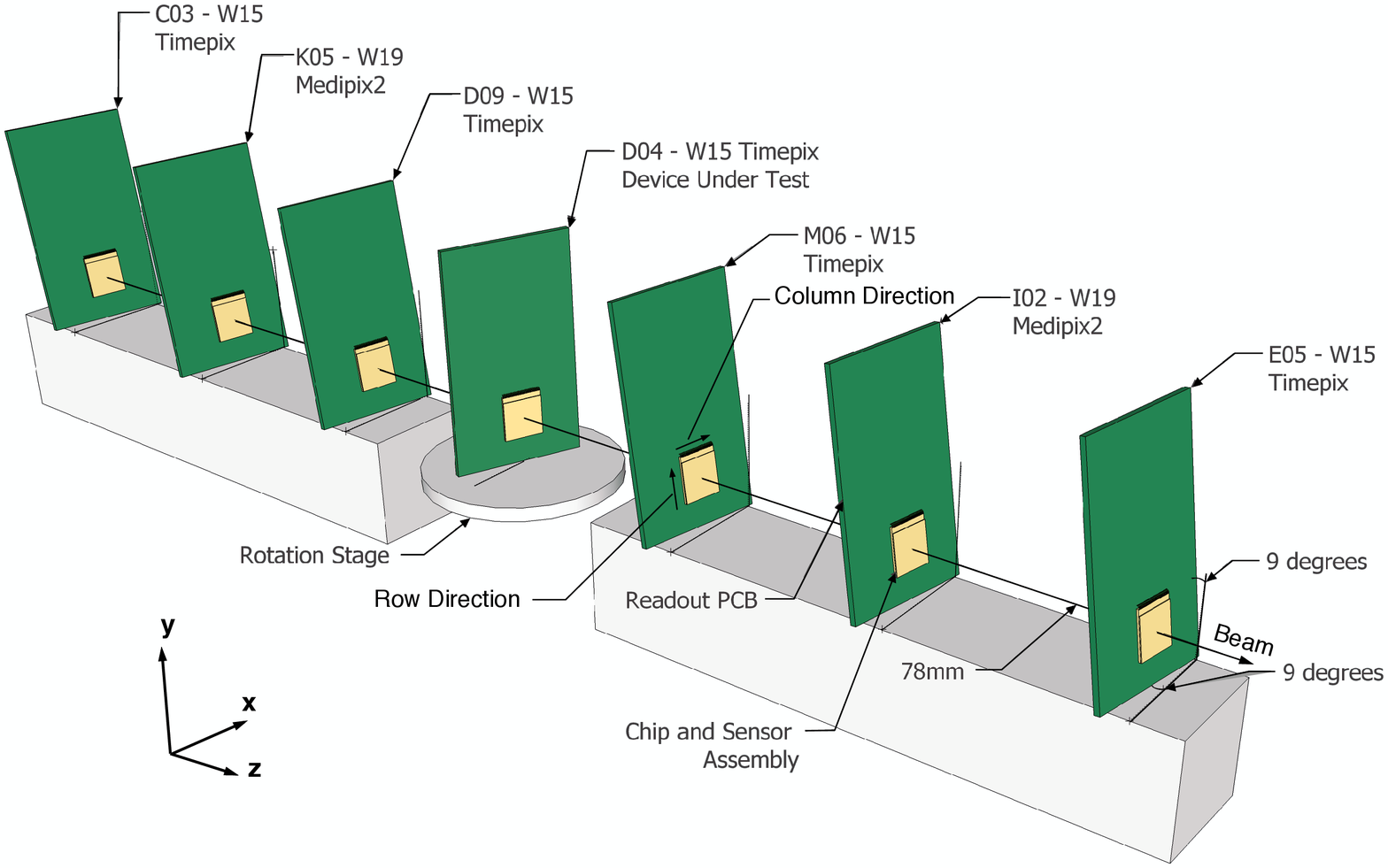}
	\end{center}
  \caption{A diagram of the pixel sensor assemblies within the telescope, showing the angled four Timepix and two Medipix2 sensors
and the Timepix DUT with its axis of rotation. The separation of the devices shown is as measured from the centers of the chips and was
enforced by the angle and the connection to the readout systems.}
  \label{fg:telescope_schem}
\end{figure}

The telescope, shown in Figure~\ref{fg:telescopephoto},
was constructed with four Timepix and two Medipix2 planes.
This allowed the easy integration of the Device Under Test (DUT), provided complete compatibility between the readout systems
and enabled an event rate of 200~Hz. The layout of the telescope
together with the global coordinate frame is shown in Figure~\ref{fg:telescope_schem}. Note the labels
identifying each chip, which will be used on certain plots in the remainder of this paper.
\epspicz{1}{2}{telescopephoto}{Photograph of the Timepix telescope used in the testbeam.}

The telescope consisted of six pixel planes, four Timepix and two Medipix2 separated from each other and the DUT by 78~mm. 
As the Medipix2 planes only provide binary information, and so produce
a lower resolution, they were sandwiched between the Timepix
sensors; the Timepix sensors form the innermost and outermost stations of the telescope arms.
A particle passing through the sensor at an angle of  $\tan{^{-1}(\frac{\textrm{Pitch}}{\textrm{Thickness}})}$
will always traverse more than one pixel. This is, therefore, an
approximation of the optimum angle of the sensor to the particle beam,
at which position resolution can be improved by weighting the charge
deposition in the pixels (centroiding).
When the sensor is at this angle the centroiding calculation incorporating the ToT information from the Timepix can be used for all recorded tracks,
as all the tracks will leave multi-hit clusters. Following this logic the telescope sensors were fixed to $9^{\circ}$ in both the
horizontal and vertical axes perpendicular to the beam line to optimise the spatial resolution that could be achieved by the telescope.
The DUT was mounted at the centre of a symmetric arrangement of chips to further increase the resolution that could be achieved.  

To allow a resolution measurement to be taken with the DUT at as many angles as possible, and to increase precision, the DUT was mounted
on high precision rotation and translation stages~\cite{stages} driven by stepper motors that allowed it to be turned and aligned remotely. 
The stages used were supplied by PI\footnote{Physik Instrumente GmbH,
  D-76228 Karlsruhe} and allowed a repeatability of 2~$\mu$m and
  50~$\mu$rad for the translation and rotation states respectively\footnote{The reference numbers of
    the components used are M403.42S and M-060.2S}.

The Medipix2 and Timepix assemblies, including the DUT, were read out using USB driven systems provided by CTU~\cite{USB} and the Pixelman
data acquisition and control software~\cite{pixelman}. These are the standard, portable, low bandwidth readout systems used in most Medipix
applications to date. Each USB unit is attached to one chip. A signal is applied simultaneously to all the USB readout units and its rising
edge triggers local shutters to individual chips. The length of the shutter is programmable in each USB unit and it was set to be the same
for each assembly. It was optimised on a run by run basis to capture between 100 and 500 tracks per frame depending on beam conditions.
The micro-controller in the USB unit introduces a delay between the trigger being received and the shutter being sent of $4.0\pm0.5$~$\mu$s. 
As shutter periods of down to 10~ms were used depending on the beam
intensity the error on the efficiency measurement introduced by this jitter should be small.
In situations where a shorter shutter is required, such as high particle flux environments, this will become a constraint on the use of the existing USB systems.

\section{Devices under test (DUTs)}
For the data taking described in this paper, two devices were
investigated in the central position of the telescope; a standard
planar sensor similar to the devices making up the telescope planes
themselves, which was used to make a thorough investigation of the
Timepix performance, and a 3D sensor, where the most interesting feature was the performance of the sensor itself in combination with the Timepix chip.

\subsection{Planar sensor}
\label{planarsensor}
The planar sensor used was one of a standard series of sensors
produced by CANBERRA\footnote{CANBERRA Industries, Semiconductor NV Belgium.} for the Medipix2 collaboration.  The
substrate is n-type, and the device has $p^+$ electrodes for hole
collection.  The substrate resistivity was 32~${\rm k}\Omega {\rm cm}$,
corresponding to 10~V full depletion voltage for the $300~ {\rm \mu m}$ thick
device.  The leakage current of the device used in this testbeam was
less than 1~nA at 40~V applied voltage. The bump bonding assembly was
carried out at VTT\footnote{ VTT Technical Research Centre of Finland, P.O. Box 1000, FI-02044 VTT, Finland} using a
Timepix chip with no inactive channels, and there is
expected to be negligible loss due to noisy pixels.  

\subsection{3D sensor}
\label{s:3dsensor}
A 3D sensor has an array of n- and p-type electrode columns passing through the
thickness of a silicon substrate, rather than being implanted on
the substrate’s surface as in a planar sensor. These electrodes are realized by a
combination of micro-machining techniques and standard
detector technologies~\cite{3DFabrication}.
By using this structure it is
possible to combine a standard substrate thickness of a few
hundred micrometers with a lateral spacing between electrodes
of a few tens of micrometers. The depletion and charge
collection distances are thereby dramatically reduced, without reducing
the sensitive thickness of the sensor. This implies that the
device has extremely fast charge collection and a low
operating voltage even after a high irradiation dose. The short
collection distance and the electric field pattern in the device
also reduce the amount trapping that takes place. These
features make 3D sensors potentially very radiation hard; in~\cite{3DIrradiation} it is shown that a $285$~$\mu$m
thick double-sided 3D sensor irradiated to $10^{16}$~1~MeV neutron equivalents/$\rm{cm}^{2}$
biased at 350~V collects 20,000 electrons. 
Hence, 3D sensors are a promising technology for inner layers of vertex detectors
at the LHC upgrades, which require operation beyond this fluence.

The double-sided 3D sensor~\cite{firstDoubleSided3D} used in the
testbeam measurements described here is a modification of the traditional same-side 3D design~\cite{3DParker}. 
In the same-side 3D devices the n- and p-type columns penetrate through the full thickness of the substrate, and the columns for both
electrode types are etched from the same side and a handle wafer, wafer-bonded to the sensor wafer, is required in the
processing. The double-sided device has the columns etched from opposite sides of the wafer for each type of column
doping. This removes the necessity to have a handling wafer and improves production yield and lowers fabrication costs. 

\begin{figure}[h]
 \begin{center}
   \includegraphics[width=0.8\textwidth]{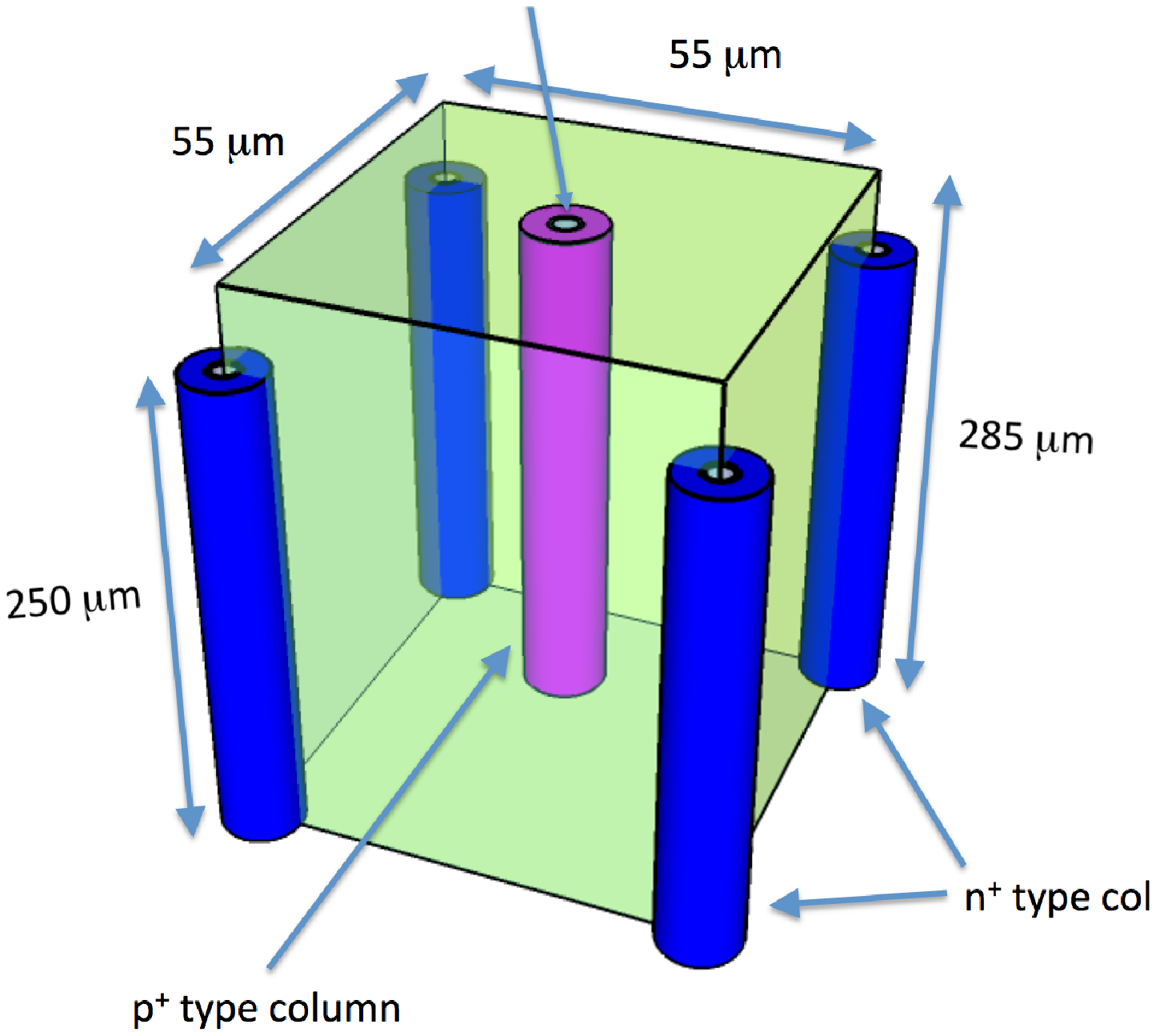}
 \end{center}
\caption{Design of a single pixel cell in an n-type bulk double-sided 3D sensor, showing the
 electrode columns partially etched from the top and bottom surfaces of the
 sensor.  The vertical scale has been compressed.}
\label{fg:3DSchematic}
\end{figure}

The double-sided 3D sensor used in this beam test was designed by the University of Glasgow and IMB-CNM and
fabricated at IMB-CNM\footnote{IMB-CNM, Centro Nacional de Microelectr{\'o}nica. Campus Universidad
Aut{\'o}noma de Barcelona. 08193 Bellaterra (Barcelona), Spain.}. 
The geometry of the sensor, with electrode columns passing through the thickness of the wafer from
the front and back sides, is illustrated in Figure~\ref{fg:3DSchematic}. The $5$~$\mu$m radius
columns are produced using deep reactive ion etching, and filled with 3~$\mu$m thickness
polysilicon around the pore walls. The columns are doped by diffusion through the polysilicon
with Boron (p-type columns) or Phosphorous (n-type columns), and the inside of the columns
passivated with Silica. The back surface is coated with metal for biasing and readout electrodes
deposited on the front-surface. Hole and electron collecting devices were fabricated. The
device used here is a hole collecting double-sided 3D n-type sensor with an n-type substrate and
p-doped columns connected to the electronic readout. The columns have a depth of $250$~$\mu$m,
whereas the substrate is $285\pm 15$~$\mu$m thick. The region between the n- and p-type columns depletes at only a few
volts. There is a lower field region in the sensor directly above a column which requires the
sensor to be biased to around 10~V to be fully active~\cite{3Dsimulation}.  The device was
solder bump-bonded by VTT
to a Timepix readout chip. A lower grade Timepix readout chip was used for this R\&D project with
some inactive pixel columns (see later). The electrical characteristics of the double sided 3D
devices measured in the laboratory are reported in~\cite{3DIrradiation}  and~\cite{firstDoubleSided3D},
and the device used here had a leakage current of $3.8$~$\mu$A at 20~V at room temperature. 

\section{Charge calibration of DUT}
\label{s:calibration}
For the data
described in this paper,
the standard threshold setting and the mean THL DAC values returned by the
equalisation scans described in Section~\ref{s:timepixchip} are shown in Table~\ref{t:THLs1}.
\begin{table}[ht]
\label{t:THLs1}
\begin{center}
\begin{tabular}{ccccc}
\hline
 & Noise Floor & Noise edge & Threshold Set & Threshold Set\\
&(DAC steps) & (DAC steps) & (DAC steps) & (electrons)\\
\hline
D04-W0015 & 460 & 438 & 400 & $\sim$1520 \\
\hline
\end{tabular}\
\caption{Planar Sensor DUT Threshold settings.  The nominal threshold in
 electrons is given by the difference between the set threshold and
 noise floor, multiplied by the expected 25.4 electrons per DAC step~\cite{LlopartCudie:1056683}.}
\end{center}
\end{table}

In order to correctly interpret the DUT data in the analyses which follow, a charge
calibration of the planar sensor DUT was performed.    
The first aim of this charge
calibration was to verify that the threshold
indeed corresponded to the nominal value
shown in Table~\ref{t:THLs1}.   The
second aim of the charge calibration was to be able to obtain the
relationship between raw
ToT values and charge deposited in each
pixel.  As explained in Section~\ref{s:timepixchip}, the response of the Timepix is in principle a linear function
of the input charge. In practice, however, there is a deviation from
this behaviour, giving a global offset for each hit pixel with a
value dependent on the threshold. In addition, there exists a strongly
non-linear behaviour for charges within about 3000~e$^-$ of the
threshold.  This behaviour was first described in~\cite{medipix-web},
where a so-called ``surrogate function" was used to parametrise the
response.  
In the case of the detection of minimum ionising particles, with thresholds similar to those used in the testbeam described here, the offset is positive, causing the raw counts to overestimate the charge deposited per pixel. Also, the non linear behaviour close to threshold affects the apparent charge sharing distributions between pixels.
In addition there is a pixel-to-pixel
variation, which the calibration can in principle detect, and correct for.
The calibration was performed with testpulses and cross checked with
data taken with a radioactive source.  The method used has not been
previously described, and so is explained in detail here.
For the 3D sensor DUT, a detailed charge calibration was not
performed, however the conclusions from the calibration of the planar
sensor DUT can be extrapolated to this device to infer the operational
threshold.

The first part of the calibration proceeded using the testpulse
function available in the Timepix and controlled via the
Pixelman interface.
The analog input of each pixel
features a built-in capacitor of $7.5$~fF~\cite{timepix}, 
and a step voltage of 1~mV
therefore corresponds to 
46.9~e$^-$ injected charge~\cite{timepix}. The voltage of the
pulse sent to the chip is generated by an analog
multiplexer\footnote{Part reference MAX4534, manufactured by Maxim
  Integrated Products, Sunnyvale, CA 94086, USA}
mounted on the board, which switches the output from a reference
voltage.  The precise relationship between the programmed voltage and
the voltage received by the chip varies according to the USB readout system linked
to the chip in question,
and must be checked by measurement at the output of the multiplexer.
For the planar sensor DUT the following relationship is obtained:
\begin{equation}
H =H_{\rm NOM}*0.972- 26.7,
\label{eq:tpvalue}
\end{equation}
where $H$ is the true test pulse height in mV, and $H_{\rm NOM}$ is
the nominal test pulse height.

In the ideal case, the calibration would now proceed by sending one testpulse per shutter to each pixel, and measuring the ToT and efficiency numbers for each testpulse height. In the case of the Timepix chip this is not possible, as the first testpulse sent gives a lower value than expected in the pixel array, an effect which is thought to be due to the settling time of the DACs in the setup used.
This effect can be washed
out by sending a large number of testpulses, however in this case the efficiency (``counting'') measurement
and the ToT measurement must be done in two separate steps, as described below.  In addition note that
it is important to send the testpulses one at a time to individual pixels within an 8 by 8 array, with a suitable
wait period between each testpulse, or the results can be misleading.  

The response was first studied in
counting mode, in which case no ToT information is provided by the chip. One thousand
test pulses were sent per frame, and the mean number of counts per pixel measured.
Three thresholds were studied, at DAC values of 380,
400, and 420, corresponding to nominal thresholds of roughly 2030,
1520, and 1020 electrons.  For each pixel, a curve can be constructed
of the number of counts (efficiency) as a function of the programmed
testpulse voltage.    Fitting an error function to the curve gives the
threshold for each pixel at the 50~$\%$ efficiency point, as well as
the noise of that pixel.  Figure~\ref{fg:efficiency_1} shows
the average efficiency of each point, with error bars indicating the
RMS spread, together with the average of the
65k curves derived and fitted in this way.    The programmed 
testpulse values are corrected to the true values using Equation~\ref{eq:tpvalue} and finally to electrons using
the 46.9e$^-$ per mV scaling given above.  In this way the average thresholds
corresponding to 420, 400, 380 DAC counts were determined to be 1030, 1550, and
2050~e$^-$ respectively.    The average pixel noise was measured to be
122 electrons, with an RMS spread of about 3 electrons.  These
calibrated threshold values can be seen to be very close to the
nominal ones. 

The CPU time
involved in performing these fits is considerable, however it is also possible
make a fast derivation of the average
thresholds by plotting the average efficiency of all pixels
a function of applied testpulse voltage and fitting with a single
error function.  This gives results within ten electrons of
the individual pixel fits, however it should be noted that in this
case the measured noise is a combination of the individual pixel noise
and the threshold spread, which contributes about thirty electrons in
quadrature to the measured spread.   These
data can also be used as an indirect cross check of the testpulse
capacitor value, due to the fact that each DAC step is expected to
correspond to 25.4~e$^-$, as reported in~\cite{LlopartCudie:1056683} from a
measurement with sources of a different Timepix chip.  The differences
in thresholds between the three measurements 20 DAC steps apart
can be seen to correspond very well to this expected value.  Finally,
note that at the lowest threshold an efficiency rise is visible at
very low values of applied testpulse voltage.  Due to the offset shown
in Equation~\ref{eq:tpvalue}, low values of applied testpulse
voltage become negative.  Each testpulse generates negative and
positive pulses in the pixel array, corresponding to the rising and
falling edge of the testpulse.  When the testpulse reverses sign, the
pixels show a response corresponding to the oppositely signed edge,
and this response is mirrored about the 26.7~mV offset given in Equation~\ref{eq:tpvalue}.
\epspicz{1}{1}{efficiency_1}{Average efficiency curves
  obtained with the counting mode calibration.  The points show the
  experimental data, with the applied testpulse voltage on the x
  axis, and the error bars indicate the
  RMS scatter.  The curves
  are the averaged curves of the individual pixel fits, as described
  in the text.}

The second part of the calibration proceeded with the chip in ToT mode, in order
to derive the relationship between the input charge and the ToT
value.  Note that in this mode the counting function is not available, so the efficiency curves described above cannot be simultaneously extracted.
We applied fifty testpulses per shutter and plotted the resulting ToT
value of each pixel, divided by fifty, as a function of applied testpulse voltage.
As described in \cite{medipix-web} the curve is well described with
the following ``surrogate function'':
\begin{equation}
{\rm 
f(x) = ax+b-\frac{c}{x-t}},
\end{equation}
where $\rm{a}$ and $\rm{b}$ represent the slope and intercept of the line describing the behavior well above
threshold, while $\rm{t}$ and $\rm{c}$ parameterize the non-linear behavior close to threshold~\cite{surrogate-function}.
In practice, electronic noise, in combination with the fact that fifty
pulses are sent per shutter, smoothes the sharpness of this
turn-on. We therefore fitted the measured calibration data with
a convolution of the surrogate function with a Gaussian whose $\sigma$ is constrained to the one extracted for
the corresponding pixel from the efficiency curves derived in
counting mode. Figure~\ref{fg:gain_convolute_2} shows the
calibration curves of the DUT  for the three thresholds
studied. Each data point is averaged over the 65k pixels.
Near the origin of the graph one can see the edge of the response for negative
pulses.   It is also possible to extract the so called
``minimum detectable charge'', corresponding to the intercept of the 
surrogate fits with the \textbf{x} axis, corrected according to 
Equation~\ref{eq:tpvalue} and converted into electrons.
This number is close to the 
threshold values quoted above, but is expected to be slightly lower due to the 
fact that many pulses per shutter are sent.   
Table~\ref{tab:fitpara}
shows the averages of the individual pixel parameters describing the fitted
surrogate functions.   The parameters take into account the correction
described in Equation~\ref{eq:tpvalue}.   
\begin{table}[hbt]
\begin{center}
\caption{Summary of the surrogate function fit parameters from the ``ToT'' testpulse scans
at three different thresholds, averaged over the individual pixel fits}\label{tab:fitpara}
\begin{tabular}{ccccccc}
\hline
THL~~ & $a$ & $b$ & $c$ & $t$ & Minimum Detectable Charge (e$^-$)\\\hline
420~~ & 0.194 & 34.4 & 195 &15.3 & 957  \\
400 ~~& 0.201 & 24.8  & 174 & 25.9 & 1480 \\
380 ~~ & 0.206& 17.6&150 & 36.4   & 1980 \\ \hline
\end{tabular}
\end{center}
\end{table}

As in the case of the counting
mode measurements, the average parameters can also be obtained in a
fast way by fitting the same function to the curve of the average ToT value as a
function of testpulse.  The numbers obtained are very close to the
numbers shown in Table~\ref{tab:fitpara}.  
Apart from the threshold behaviour, the main characteristic exhibited by these fits is a positive offset of the linear part of
the curve, which results in a strong difference seen in the pixel spectra depending on the number of pixels per cluster. 
The calibration data can be used to calculate the detected charge in
units of electrons, given a measured ToT, using the following
``inverse surrogate function'':
\begin{equation}
{\rm 
q_{in}(e^-)=46.9*\frac{t*a+ToT-b+\sqrt{(b+t*a-ToT)^2+4*a*c}}{2*a}.
}
\end{equation}
which simplifies in the linear regime (above the threshold region) to:
\begin{equation}
{\rm 
q_{in}(e^-)=46.9*\frac{ToT-b}{a}.
}
\end{equation}
The parameters used can be those of the individual pixels, or the
average fitted curves.  In the latter case, there is an additional
spread of less than 5$\%$, for charge deposits around the value
expected from a minimum ionising particle.

As a cross check of the studies described above, the response of the DUT to a $^{55}$Fe source was measured,
which features a 5.9~keV X-ray line, producing a charge deposition of 1640~e$^-$ in Silicon. The DUT was exposed
to this source and the response measured with the device configured in counting mode is shown in Figure ~\ref{fg:ironformarina}.
The main curve shows the measured efficiency curve, while the insert shows its derivative. The latter curve exhibits a
clear peak at a threshold of 403 DAC counts. From this study it can be inferred
that a threshold of 400 corresponds to 1560~e$^-$,
consistent with the result obtained from the test pulse calibration. 
\epspicz{1}{1}{gain_convolute_2}{Average test pulse response of the DUT in ToT mode, for three different thresholds. The points represent show the
pulse response averaged over the 65k pixels in the DUT, as a function of the programmed testpulse voltage, with error bars
indicating the RMS scatter.  The curves represent averages of the individual pixel fits to the surrogate function, as described
in the text.  The I$_{\textrm{krum}}$  setting, which determines the conversion gain, was 5.}
\epspicz{1}{1}{ironformarina}{Average efficiency curve obtained by operating the Timepix ASIC in counting mode exposed to a $^{55}$Fe source in counting mode.}
\section{Treatment of telescope data}
\label{s:datatreatment}
\subsection{Clustering algorithm}
\label{s:clustering}
\subsubsection{Cluster finding}
A cluster is defined as a group of adjacent hits surrounded by empty pixels. For two hits to be considered adjacent, they must simply
have a difference of $1$ in their \textbf{x} or \textbf{y}
positions.
The cluster maker works by making a cluster out of the first hit on the sensor and recursively searching for adjacent hits; each time
an adjacent hit is found it is added to the cluster and the search for adjacent hits repeated. The search stops when no
further adjacent hits can be added to the cluster. 
The position of the cluster in local coordinates is assigned by making
a charge weighted average of the individual pixel positions in the \textbf{row}
and \textbf{column} directions.


\subsubsection{Telescope cluster properties}
The distribution of ToT values in the individual pixels, and the
distribution summed over the clusters, are shown in
Figure~\ref{fg:simpleclustercleaning} for a typical telescope plane.
There is a small amount of noise below the Landau-like cluster peak
which occurs due to interactions from showers in the North Area pion beam; if
it is required that clusters are within 100~$\mu$m of a reconstructed track (where that device is
excluded from the track fit) this noise is removed, as can be seen on
the right hand plot. 


In Figure~\ref{fg:landaus} the total cluster charge of 
one, two, three, and four pixel clusters is shown for a sample telescope plane, angled at
$9^\circ$ in the horizontal and perpendicular directions. 
It can be seen that the ToT of the clusters increases incrementally with each additional pixel, with most
probable values for this plane of 119,145,168, and 203 ToT counts for the
one, two, three, and four pixel clusters respectively.
This effect is due to the behaviour of the Timepix chip which gives a
ToT value which is proportional to the charge plus an offset, and
a turn-on behaviour in the threshold region, as discussed in
Section~\ref{s:calibration}.  A more detailed analysis of the
cluster characteristics was performed for the DUT and is described in
Section~\ref{s:analysisofdeviceundertest}.
For the purposes of the telescope performance
this effect is not expected to have a large impact on the pointing
resolution, and for the analyses described in this paper it remains
uncorrected in the telescope data.
For the purposes of the tracking, the candidate telescope clusters have to pass only one cut, namely that the ToT is greater than 80 and less than
350 counts. 

\epspicz{1}{1}{simpleclustercleaning}{The signals in individual pixels (dashed), and the signals in clusters (solid)
after the clustering algorithm, shown on the left plot for a sample telescope plane.
There is a small amount of noise below the Landau peak around 50. If the clusters are required to be within
100~$\mu$m of a track this noise is removed, as shown on the right
hand plot. The curves are normalised to unit area.}
\epspicz{1}{1}{landaus}{ToT of clusters from a sample plane in
  the telescope. The curves are normalised to each other for
  illustrative purposes.  The difference in the MPV between each
  population is due to the positive offset introduced in each pixel by
  the Timepix gain curve.
The relative population of one, two, three, and four pixel clusters is dependent on the device angle and is discussed in Section~\ref{s:analysisofdeviceundertest}. The
solid black vertical lines show the positions of the applied cuts.}
\subsection{Tracking}
\label{s:patternrecognition}
The track reconstruction proceeds in two stages, a pattern recognition
followed by a track fit.  A global event cut of 5000 hits
in the whole dataset is applied to reject a very small sample of
saturated events.
A reference sensor is chosen towards the centre of the telescope and the global \textbf{x} and \textbf{y} of clusters in the
other planes are compared with this. The resulting distribution for all clusters is shown in Figure~\ref{fg:inspectglobalalignment}
for an adjacent plane and for the most remote plane.  It can be seen that even in the most remote plane, the track related clusters
are all contained within a window of $\pm 100$~$\mu$m. Clusters are
selected within this window, and it is required that every track contains a cluster from every telescope plane.
As the occupancy is around 0.2$\%$ the background picked up by this procedure is negligible.
The clusters are then fitted with a straight line track fit, which takes into account the full rotation matrix of
each pixel plane and is able to assign individual errors to each plane.  For the purposes of this paper one error is
applied to the Timepix planes and another to the Medipix2 planes.  Most data were taken with a fairly diffuse beam
covering the full surface of each sensor. Due to alignment effects there is an acceptance of about $85\%$ in the
telescope. In a typical run of 1000 frames, there are about 100k clusters reconstructed in each plane, and approximately
64k tracks. Note that due to time alignment effects with the edges of
the SPS spills some frames are empty.
Typically the number of tracks reconstructed per spill was about 750.

\epspicz{1}{1}{inspectglobalalignment}{The global \textbf{x} and \textbf{y} of clusters in a given telescope plane compared to the \textbf{x} and \textbf{y} of clusters
in the reference plane. The left hand plots show these for a telescope plane next to the reference, the right hand plots show them for
the telescope plane farthest from the reference.}
\subsection{Telescope alignment}
\label{alignMM}
The telescope was not surveyed prior to the testbeam due to the fact that the chips were not precisely placed on
the chipboards, which were not designed for this purpose. Instead, a software alignment was used,
with the only input being the rough \textbf{z} positions and inclinations of the planes.  The coordinate system is the
right handed one introduced in Section~\ref{s:telescope}: \textbf{x} is perpendicular to the beamline, parallel
to the floor, \textbf{y} is perpendicular to the beamline, perpendicular to the floor, and \textbf{z} is along the beamline.
The free parameters adjusted for the alignment are then
(\textbf{x},\textbf{y},\textbf{z}, $\theta_{\rm{\textbf{x}}}$, $\theta_{\rm{\textbf{y}}}$ {\rm{and}} $\theta_{\rm{\textbf{z}}}$) where
$\theta_{\rm{\textbf{x},\textbf{y},\textbf{z}}}$ are rotations about the \textbf{x},\textbf{y},\textbf{z} axes.  Physically the planes of the telescope are positioned
with the sensors away from the beam, rotated about the \textbf{x} and \textbf{y} axes by $9^\circ$, e.g. (0,0,\textbf{z},$9^\circ$,$9^\circ$,0).
A multistep offline procedure is used to refine the sensor positions.  The first step roughly aligns the planes
by minimising the global \textbf{x} and \textbf{y} positions of the clusters in each plane relative to a reference plane; allowing
only \textbf{x}, \textbf{y} and $\theta_{\rm{\textbf{z}}}$ to vary.  With the sensors now roughly aligned, tracks are fitted to the clusters,
only accepting tracks with a slope less than 0.005 radians.  A $\chi^2$ is formed for each track from the sum of
the residuals divided by the expected error in each plane.  The minimisation proceeds by varying the positions of
all telescope planes individually in all coordinates except \textbf{z}, and iterating three times.  In a final step, the procedure is repeated but
the telescope planes are also allowed to move in \textbf{z}.  New tracks are produced
and extrapolated to the DUT position.  The DUT is always excluded from the pattern recognition and the track fitting,
to ensure a completely independent measurement.  In the final step of the procedure the DUT is aligned by selecting
all clusters within a cut, typically 100~$\mu$m, and adjusting all 6 free parameters of the DUT.  The quality
of the alignment can be probed by plotting the means of the resulting unbiased residuals at the DUT as a function
of various parameters.  The most sensitive parameters are found to be the orthogonal coordinate of the track to the
residual plotted.  The distribution is shown in Figure~\ref{fg:alignmentquality}
for two sample runs with the DUT at angles of -2$^\circ$ and 8$^\circ$: these distributions are
typical.  The variation of the mean of the residual across the width of the chip is below $\pm$1~$\mu$m; for
most of the residuals measured the effect is negligible, and for the most precise measurement angles of the DUT
the resolutions extracted in Section~\ref{s:analysisofdeviceundertest} are approximately 0.2~$\mu$m
worse than what could be achieved with an optimal alignment.

\epspicz{1}{1}{alignmentquality}{An illustration of the quality of the alignment for two different runs.  The unbiased residual at
the DUT is shown, where the DUT is excluded from the pattern recognition and from the track fitting.
The variation of the mean of the residuals across the length and height of the chip is illustrated, for
a run close to perpendicular, where the DUT resolution was close to 9~$\mu$m, and a run close to the
optimal angle, where the DUT resolution was close to 4~$\mu$m.}
\subsection{Track pointing precision}
\label{s:pointingresolution}
After the pattern recognition a straight line track fit is performed on the clusters selected in the telescope.
For the majority of the analyses, it is required that each track has a cluster from each of the six planes.
Due to the fact that the planes equipped with Medipix2 sensors are expected to have lower resolution than the Timepix
planes, the track fit takes into account the errors in order to weight the clusters appropriately in the fit.
The distributions of the biased residuals are studied to extract the resolution of the individual planes and the
pointing resolution of the telescope at the position of the DUT. Simplifying the expression such that the telescope
planes are arranged symetrically around $\rm{\textbf{z}}=0$, it can be
shown, following the formalism developed in~\cite{Papadelis:1186697}, that the predicted position of the track at plane $n$ is given by
\begin{equation}
{\rm {\rm y_{pred(n)}} = \displaystyle\sum_i y_i \cdot A(z_{n,i})}, 
\end{equation}
where
\begin{equation}
{\rm A(z_{n,i}) = \frac {\displaystyle\sum_j \frac{(z_j)^2 + z_n \cdot z_i}{ (\sigma_j)^2 \cdot (\sigma_i)^2}}{\displaystyle\sum_j \frac{1}{\sigma_j^2} \displaystyle\sum_j \frac{z_j^2}{\sigma_j^2}}},
\end{equation}
and $\rm z_n$ is the \textbf{z} position at plane $n$, and ${\rm \sigma_n}$ the error assigned to plane $n$.
The expected track pointing error is then given by: 
\begin{equation}
{\rm {\rm \sigma_{pred(z_n)}} = \sqrt{\displaystyle\sum_i (\sigma_i \cdot A(z_{n,i}))^2}}.
\end{equation}

The quality of the residuals in the telescope is illustrated in Figure~\ref{fg:telescoperesidualsforpaper},
which shows an example of the fitted residuals in the \textbf{x} coordinate for the six
telescope planes. It can be seen that the shapes are
quite gaussian.   The biased residuals extracted in each plane in
\textbf{x} and \textbf{y} for a typical run are shown in
table~\ref{t:biasedresiduals}.
The distribution of the biased residuals can be used together with the formulas above to extract
the expected pointing resolution at the centre of the telescope.

\begin{table}[ht]
\begin{center}
\begin{tabular}{lcccccc}
\hline
Device name & C03-W15 & K05-W19 & D09-W15 & M06-W15 & I02-W19 &
E05-W15 \\
\hline
$\sigma_{\textbf{x}}$  (${\rm \mu m}$) & 3.3 & 9.4 & 3.9 & 3.8 & 9.3
& 3.2 \\
$\sigma_{\textbf{y}}$  (${\rm \mu m}$) & 3.4 & 9.4 & 4.0 & 3.8 & 9.2
& 3.2 \\
\hline
\end{tabular}
\caption{Biased residuals measured in the telescope for a typical
  run.  Variations between runs are of the order of less than 0.1~$\mu$m.}
\label{t:biasedresiduals}
\end{center}
\end{table}

There is an additional error introduced by multiple scattering, although the impact is small enough that
it is still acceptable to perform a straight line fit.  Each station contains roughly 2$\%$ in radiation length of material,
made up of the 700~$\mu$m thick ASIC, the 300~$\mu$m thick sensor, 1.6~mm of epoxy in the PCB and a
total of 87~$\mu$m of copper layers in the PCB.  The effect of this is assessed by simulating the setup, applying an
extra scattering to the points, and performing a straight line fit.  The result of the simulation is shown in
Figure~\ref{fg:simulationforpaper}.  The simulation gives a reasonable fit to the data, and predicts that the pointing
resolution at the position of the DUT is 2.3~$\mu$m. The dominant uncertainty on this number comes from the
fact that the data are not a perfect match to the simulation. This was investigated by varying the resolution of the
telescope until the simulation matched perfectly the internal and external planes.  The biggest deviation induced in
the pointing resolution was 0.1~$\mu$m. Additional uncertainties coming from remaning misalignments and the
lack of perfect knowledge of the material of each plane are negligible in comparison. 
\epspicz{1}{1}{telescoperesidualsforpaper}{Fitted residuals in the \textbf{x} coordinate for the six telescope planes. The middle column shows the
residuals in the two Medipix2 sensors.  Each plane is angled at
$9^\circ$ in the horizontal and perpendicular direction, and all
telescope planes are included in the track fit, with weighted errors.}
\epspicz{1}{2}{simulationforpaper}{The simulation of the straight line track fit in telescope planes 1-3 (downstream of the device under test)
and planes 5-7 (upstream of the device under test) shown as a solid line, and compared to the biased residuals measured
in the data for the \textbf{x} (open circles) and \textbf{y} (open squares) projections.}
\section{Analysis of the DUT}
\label{s:analysisofdeviceundertest}
\subsection{Description of datasets}
In total about 200 runs were taken over a two week data taking period,
each lasting about 50~minutes.  The majority of the runs had 1000
frames, with the shutter open for 
0.01~s, and a mean number of tracks of approximately 70 per frame.
Nominal settings were defined, with a threshold DAC of 400 and Ikrum 5,  
and in general one parameter was varied at a time during each investigation.
The beam intensity and the shutter open time were varied for 
specialised investigation of particular effects, such as alignment or
time-walk.   After major telescope interventions, and before taking any 
data set where the DUT was nominally
perpendicular to the beam, a quick ``angle scan'' was taken.  The 
cluster widths were analysed online in a similar way to that described in
Section~\ref{s:perpendicularpoint} to ensure
that the DUT was indeed close to perpendicular.
The principal datasets taken with the planar sensor:
\begin{itemize}
\item{\textbf{Angle scans:}
Using the motion stage the DUT was rotated about the \textbf{y} axis, causing
the clusters to spread in the \textbf{x} direction along the pixel columns.
Approximately 30 different angles were taken, concentrating on the 
region between zero and the optimal angle of around 9$^\circ$, which
is most relevant for LHCb, increasing to a largest angle of 18$^\circ$, and including negative angles to investigate the symmetry
around the perpendicular point. As a final step the DUT was rotated
by 45$^\circ$ about the \textbf{z} axis, and a further angle scan performed, in
order to investigate the resolution of the sensor between pixels
offset by one in both the column and row direction, with a constant
telescope extrapolation precision.}
\item{\textbf{Shaping scans:}
The I$_{\textrm{krum}}$  parameter was varied from 5 up to 80 in order to investigate
the effect of the shaping time on the cluster landau distributions and 
the DUT resolution.}
\item{\textbf{Threshold scans:}
The threshold DAC was varied from a lowest value of around 750 electrons through
to the highest possible value of around 6000 electrons. These data were used to
investigate the pulse shape and the DUT efficiency.  A similar scan
was performed with the beam off, to investigate the noise as a
function of threshold DAC.}
\item{\textbf{HV scans:} 
The HV was varied between 5 and 200~V to investigate the cluster
characteristics resolution of the DUT.  These scans were performed at
three different angles: $0^{\circ}$, $10^{\circ}$, and $18^{\circ}$.}
\item{\textbf{Time-walk:}
In order to investigate the time-walk the shutter open time was 
decreased to 250~$\mu$s and long data runs were acquired with the Timepix in ToA mode.}
\item{\textbf{Specialised Runs:}
A number of specialised runs were taken to cross check various
effects.
These included runs with electrons, runs with the DUT in counting mode
to check the binary resolution performance of the sensor and sparse
data runs for fine tuning the alignment.}
\end{itemize}

For the 3D sensor a subset of these datasets were taken, principally
angle scans in the normal direction, and data at three different bias
voltages. 
\subsection{Establishing the orientation of the DUT in the beam}
\label{s:perpendicularpoint}
Although the rotation stage can move and reproduce the angle of the DUT with very high accuracy 
(see Section~\ref{s:telescope}), there is an uncertainty on the absolute calibration of the angle
due to the uncertainty of placing the entire telescope perpendicularly in the beam line.  
Even though the translations and rotations of the planes relative to one another can be determined very 
precisely with the alignment, the overall angle with respect to the beam remains a weakly constrained parameter.
For this reason a wide range of angles were scanned on both sides of the nominal perpendicular position of the DUT.
Cluster characteristics independent of the tracking and alignment such as row width, column width, and fraction of 1 pixel 
clusters were then used to determine by symmetry arguments the position of the true perpendicular point relative 
to the nominal perpendicular point of the rotation stage. In order
to improve the accuracy of these measurements the cluster cleaning
described in Section~\ref{s:clustering} was applied.
\epspicz{1}{1}{RatioFractionAngle}{The left plot shows the
  distribution of the ratio of row width to column width in the $55 \times 55$~$\mu$m$^2$ planar sensor DUT; right plot shows percentage of  
various sizes of clusters as a function of nominal angle.}
\begin{table}[ht]
\begin{center}
\begin{tabular}{cc}\hline
  Scan & Minimum Angle\\ \hline
\hline
Cluster Size   & -0.205\\
\hline
Cluster Row Width  & -0.306\\
\hline
Fraction of 1-pixel Cluster & -0.273\\
\hline
Ratio (Row/Column) & -0.366\\
\hline
\end{tabular}
\caption{The DUT angle at which each quantity is minimized.}
\label{t:MinAngles}
\end{center}
\end{table}
Distributions of these quantities as a function of nominal angle are 
shown in Figure~\ref{fg:RatioFractionAngle}.
The plots show the ratio of the row width of the clusters (i.e. size in rows) to their 
column width (i.e. size in columns), and the fraction of various sizes of clusters as a function of angle. The plots show clear 
minima and maxima around the perpendicular position, and by fitting with polynomials the true position of the perpendicular can 
be extracted.The fits are performed in the central region of $\pm 5^{\circ}$, and 
the results of the fits are shown in Table~\ref{t:MinAngles}.  The dispersion of the beam was measured to be $0.007^{\circ}$ in the 
direction of rotation in the stage, so adds a negligible uncertainty to the determination of the DUT angle.  
From these studies, the angle of the beam relative to
the nominal perpendicular point of the rotation stage was
\begin{equation}
  {\rm {\rm \theta_{\rm{true~angle}}} = \theta_{\rm{nominal~angle}} -0.29^{\circ} \pm 0.06^{\circ} },
\end{equation}
and this correction is applied in subsequent plots.
The error on this angle comes from the 
errors on the polynomial functions fitted to Figure~\ref{fg:RatioFractionAngle}, in which
the errors on the individual measured points are around the permille level and therefore negligible. 
Essentially the entire error comes from the limited number of angles for which the measurement was performed.

\subsection{Extracting the resolution of DUT}
\label{extractingresolution}

In the sections which follow the resolution of the DUT is quoted for
varying conditions (track angle, HV).  The resolution is always
extracted in the same manner.  The tracks are fitted through the
telescope, as described in Sections~\ref{s:patternrecognition} and
Section~\ref{s:pointingresolution} with the DUT excluded from the
pattern recognition and from the track fit. The tracks are
extrapolated to the plane of the DUT and a global residual formed
between the tracks and all clusters in the DUT.
In order to investigate the resolution in \textbf{x}, clusters with a
residual of less than 100~$\mu$m in \textbf{y} are selected, and vice versa.
The resulting histograms are fitted with single Gaussians.  The
resolution is extracted by subtracting in quadrature the 2.3~$\mu$m
contribution from the track pointing precision.  
The error on the precision was estimated by dividing the datasets into two and comparing the resolution,
by varying the binning of the histograms, and by varying the cut on
the \textbf{y} of the cluster.
In addition there is slightly larger error for the perpendicular fits,
where the histograms become less Gaussian due to the more binary
nature of the resolution.
It should be noted that the 0.1~$\mu$m error on the track pointing
resolution is not 
the dominant contribution to the error on the resolution measurements, which are 
in all cases 4~${\rm \mu m}$ or above.
\section{Cluster characteristics of the sensors}
\label{s:clusterchars}
Some basic cluster distributions are shown in Figure~\ref{fg:clustercharacteristics}
for tracks in the DUT at perpendicular incidence and for a bias voltage of 100~V in the DUT.
Most of the clusters are formed from four pixels or less with a tail to larger clusters.
The distribution of charge in the individual and in clustered pixels is shown before and after
the charge calibration is applied.  The sharp peak in the individual pixel charge as well as the
importance of pixels with a low amount of charge is apparent.  The track intercept point for
associated clusters are shown as a function of position within the pixel cell for one, two, three, and four
pixel clusters in Figure~\ref{fg:clusterpositions}.  The four pixel clusters, which are slightly
favoured over three pixel clusters, occupy the corners of the pixel cell, with the three pixel clusters
slightly inset.  
\epspicz{4}{5}{clustercharacteristics}{Cluster characteristics
  in the 300${\rm \mu m}$ thick planar sensor DUT, at perpendicular
  incidence.  From top to bottom:
  Number of pixels per cluster, charge in individual and clustered
  pixels before pixel calibration, charge in individual and clustered
  pixels after charge calibration.  The individual pixel charge has a
  peak close to the expected minimum ionising particle deposited
  charge, corresponding to one pixel cluster, and a long tail
  extending to the threshold cut-off corresponding to clusters with a
  large amout of charge sharing.}
\epspicz{1}{1}{clusterpositions}{Cluster position
  distributions within the pixel cell. From left to right, the track
  incident point is shown in the case of associated one, two, three, and four pixel clusters.}
\subsection{Planar sensor Landau distributions}
\label{planarsensorlandau}
The extensive literature related to predictions of energy loss in thin silicon
detector is reviewed by Bichsel in \cite{bichsel}. It is well known that to
describe the width of the observed straggling function the effects of
atomic binding must be included. Several methods have been proposed
to do this. One approach is to make a n-fold convolution of the single collision
cross-section derived from data on optical absorption and other
measurements. An example of such an approach is discussed in detail in
\cite{bichsel}. No analytic form for this function exists. However, in
a subsequent paper Bichsel has provided tabulated data for the
cumulative probability distribution as a function of $\beta \gamma$
which allows the expected distribution to be generated using a Monte
Carlo simulation.

An alternative approach are  the so called 'mixed' methods. In these a
straggling function, obtained from a simple model of the
collision spectrum, is convolved with a distribution that accounts for
the atomic binding \cite{blunck}, \cite{hall},  \cite{hancock}. The simplest example of
this method is to convolve a Landau distribution with a Gaussian. The
resulting function is sometimes referred to as the Blunck-Leisegang
function in the literature. The Landau MPV and scale $\xi$ are constrained by the theoretical relation
\begin{equation}
MPV (keV) = \xi [12.325+\ln{\frac{\xi}{I}}],
\label{eqMPV}
\end{equation}
where
\begin{equation}
\xi = 0.017825 \times t\ (\mu{\rm m}),
\label{eqxi}
\end{equation}
and $t$ is the detector thickness in micron \cite{bichsel}. These formulae include
density effects which lead to an $8\%$ increase in the MPV for highly
relativistic particles compared to a MIP of 77.6~keV. The $\sigma$ of the
convolution Gaussian is predicted to be \cite{blunck}, \cite{hancock}
\begin{equation}
\sigma_{B} = 5.72 keV.
\label{eqsigma}
\end{equation}
 Bichsel \cite{bichsel} has questioned the
validity of this approach and shown that depending on the assumptions
made in the calculation values of $\sigma_{B}$ up to a factor larger
can be obtained. To obtain good agreement between the Blunck-Leisegang and
Bichsel functions $\sigma_{B}$ should be set to 7.3~keV.

In Fig.~\ref{fig:bichs} the charge collection distribution for all cluster sizes is compared to the
results of three fits:
\begin{figure}[htb]
\centering
\label{fig:bichs}
\includegraphics[width=0.7\linewidth]{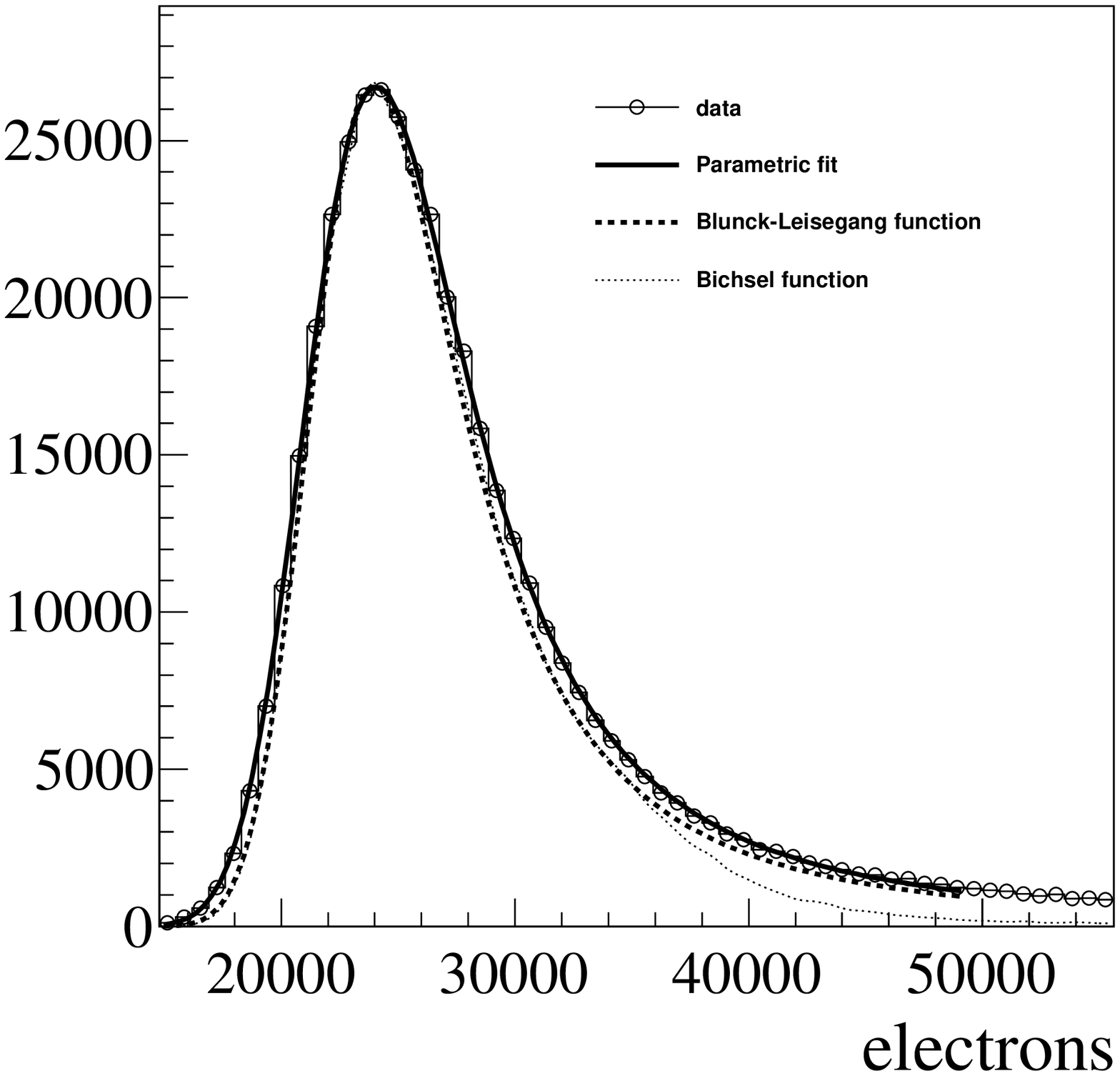}
\caption{Distribution of collected charge after charge calibration. The results of the fits
  described in the text are superimposed.}
\end{figure}
\begin{itemize}
\item A fit to the Bichsel function described above. Since the model
  itself has no free parameters only the overall normalization is left
  free in the fit. Since the tail of the distribution is expected to
  be dominated by the effect of $\delta$-rays and not well described
  by theory it is chosen to normalize to the height of the distribution.
\item A fit to a Landau convolved with a Gaussian with the parameters MPV, $\xi$
  and $\sigma_B$ fixed to the values given by Equation \ref{eqMPV}  - \ref{eqsigma}. As in the fit to the Bichsel function only the
  normalization is left free.
\item A parametric fit to a Landau convolved with a Gaussian with all
  parameters left free. The resulting fitted parameters, compared with the predicted values from
  Equations~\ref{eqMPV}~to~\ref{eqsigma} are displayed in Table~\ref{tab:landau}.
\end{itemize}
Additional broadening effects due to electronic noise and
imperfections in the gain calibration are small and can be ignored. To
convert the theoretical estimates from keV to electrons a value of
3.62~eV for the energy necessary to create an electron-hole pair in Si
\cite{canali} was used. It can be seen that though the Bichsel
function describes the shape of the core of the distribution well it
underestimates the MPV by around $3 \%$. The tail of the distribution
is relatively poorly described. On the other hand the Blunck-Leisegang
function agrees well in terms of the MPV but underestimates the observed
width of the distribution. To describe properly the $\sigma_{B}$
of the Gaussian needs to be increased from 1580 to 1830
electrons. This is in agreement with the value for $\sigma_{B}$
proposed in \cite{bichsel}.

\begin{table}
\begin{center}
\caption{Predicted and fitted Landau parameters.}\label{tab:landau}
\begin{tabular}{ccc}
\hline
Parameter& Predicted value [$e^{-}$]  & Fitted [$e^{-}$] \\ \hline
MPV & 23400 &   23410 $\pm 10$ \\ 
$\xi$ & 1690 &  $1820 \pm 10 $  \\
$\sigma_{B}$ & 1580  & $1830 \pm 10$  \\ \hline
\end{tabular}
\end{center}
\end{table}

\begin{figure}[t]
\centering
\label{fig:clustersizefit}
\includegraphics[width=0.45\linewidth]{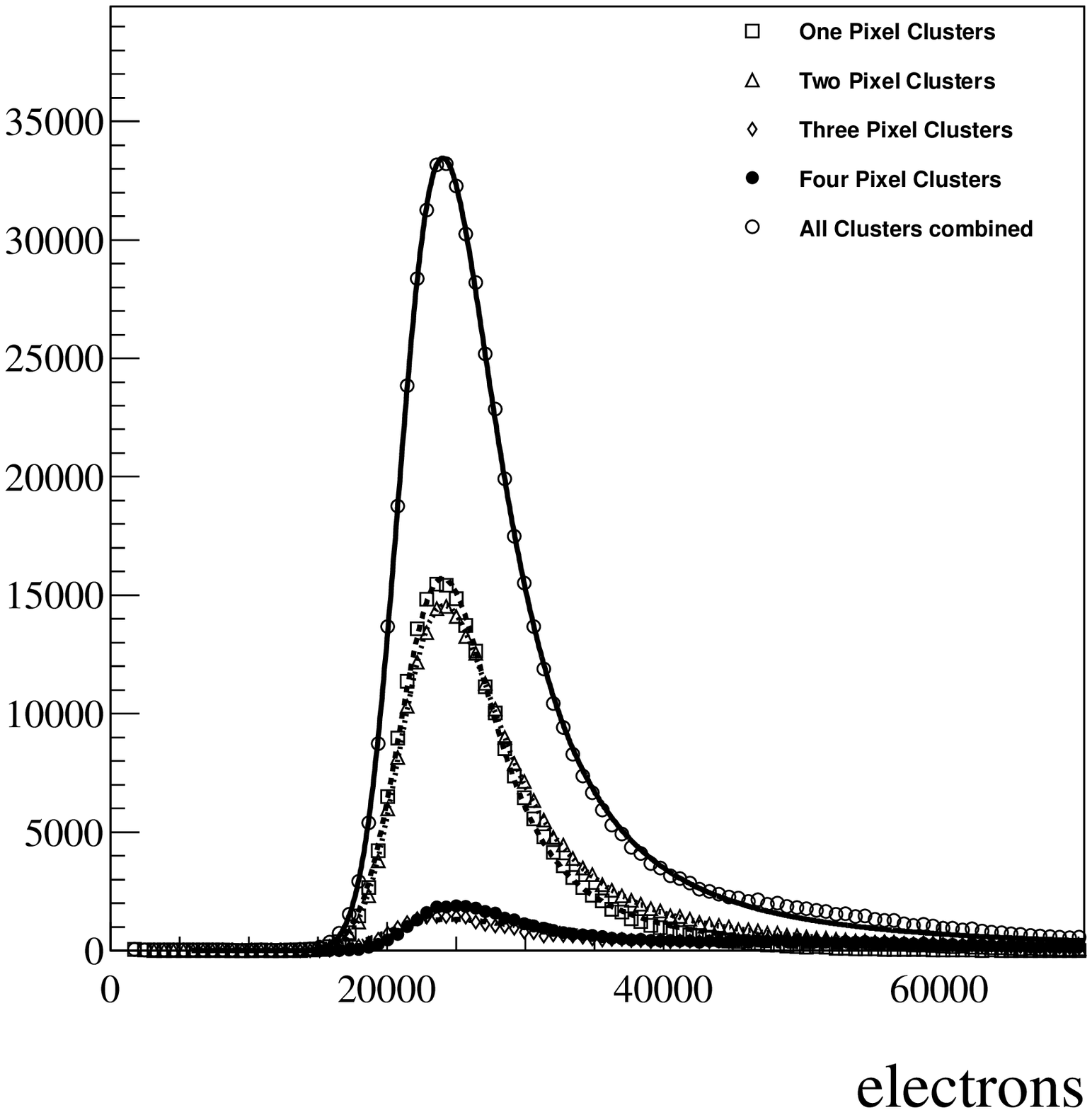}
\includegraphics[width=0.45\linewidth]{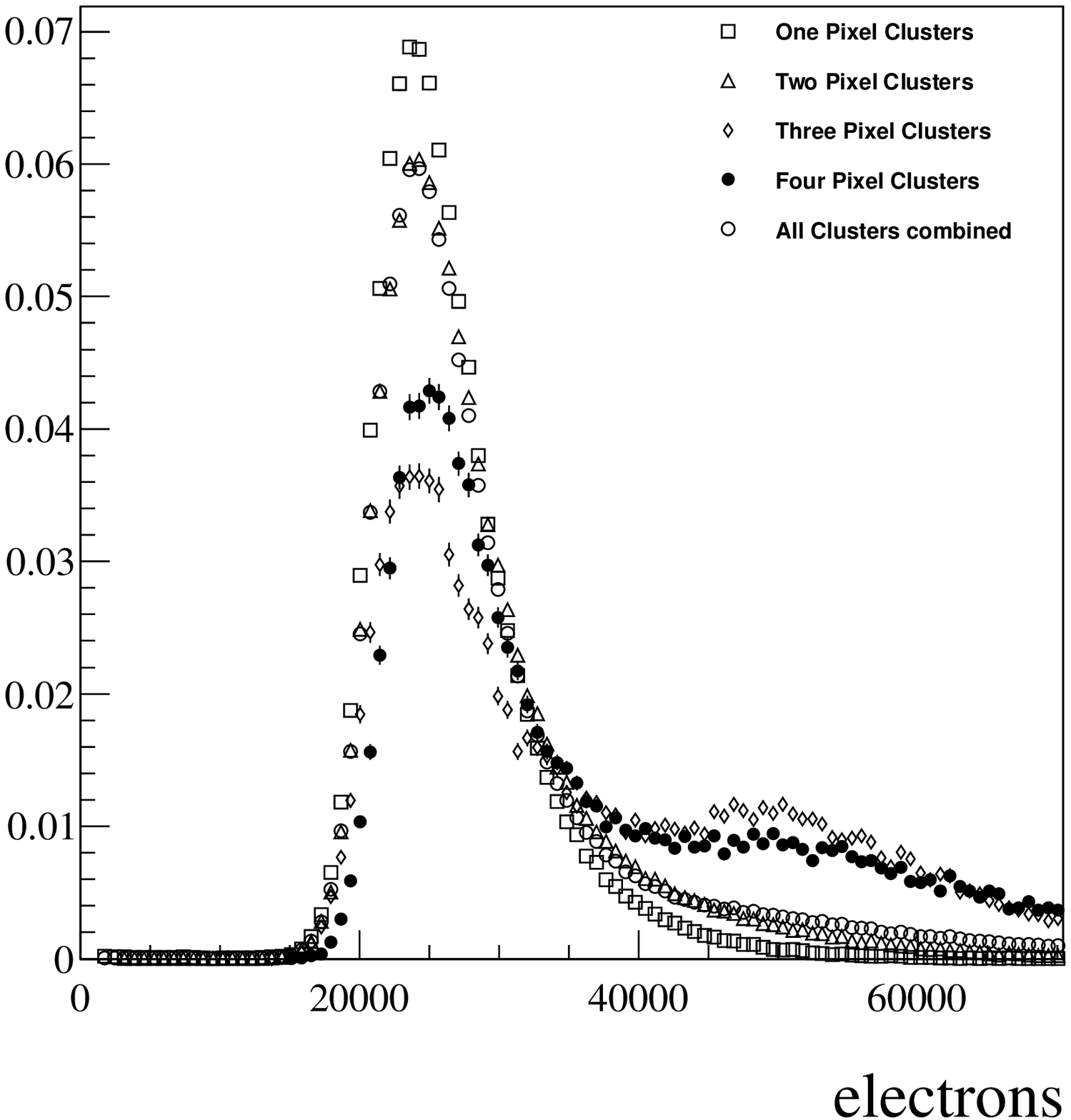}
\caption{Distribution of collected charge after charge calibration for different cluster
  sizes. The results of fits to the parametric form described in the
  text are described. The right hand plot shows the same distributions
  but normalized to unit area, and with a zoomed x-axis.}
\end{figure}

Studies have also been performed dividing the data according to the
cluster size. Fits to the parametric form described above are shown in
Fig~\ref{fig:clustersizefit}. The MPV returned by the fit are given in
Table~\ref{tab:clussizetab}. It can be seen that the MPV increases slightly with
increasing cluster size. This is most likely explained by the relatively high readout threshold DAC of $1500
e^-$ which was applied and means that some charge is lost. This
hypothesis is supported by the fact that if the track information is
used to select one-strip clusters where the impact point point is predicted to be
close to the centre of the pixel, the deposited charge increases from
24070 to $24479 e^-$ in better agreement with the numbers seen for
multistrip clusters.

\begin{table}
\begin{center}
\caption{Deposited charge for different cluster sizes.}.\label{tab:clussizetab}
\begin{tabular}{cc}
\hline
Cluster size & MPV [$e^{-}$] \\ \hline
1 & 24029  \\ 
2 & 24145   \\ 
3 &  24214 \\ 
4 &  24975\\ \hline
\end{tabular}
\end{center}
\end{table}

In addition, as the cluster size increases the distribution of the
deposited charge is observed to broaden. This can be explained by 
the fact that multi-strip clusters are more likely to occur if a high
energy $\delta$-ray is produced.

Finally, it should be noted that if an averaged calibration is applied
to the pixels instead of the individual calibration, there is no significant change to the results.
\subsection{Double sided 3D sensor Landau Distributions}
\begin{figure}[htb]
\centering
\label{fig:3dlandau}
\includegraphics[width=0.45\linewidth]{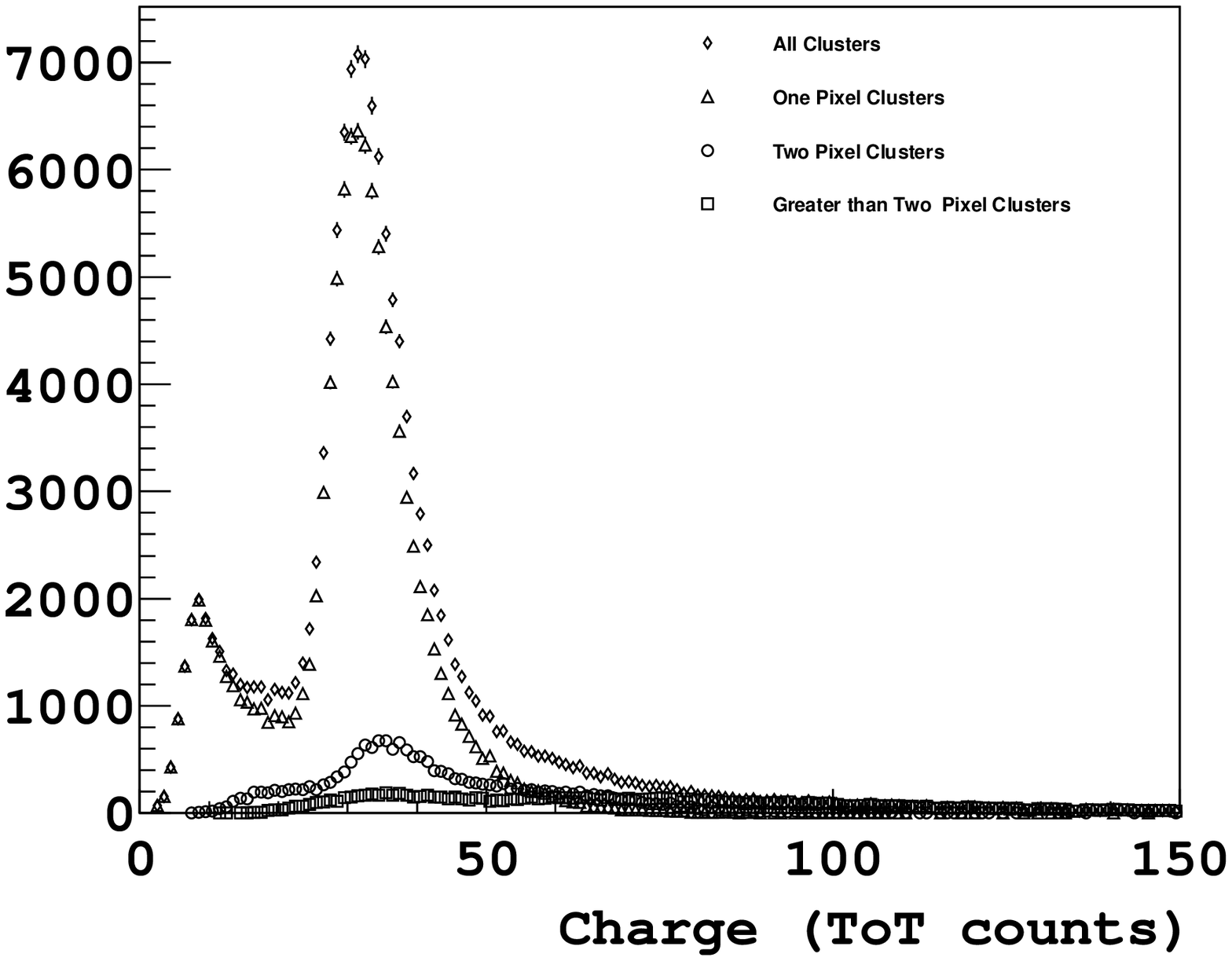}
\includegraphics[width=0.45\linewidth]{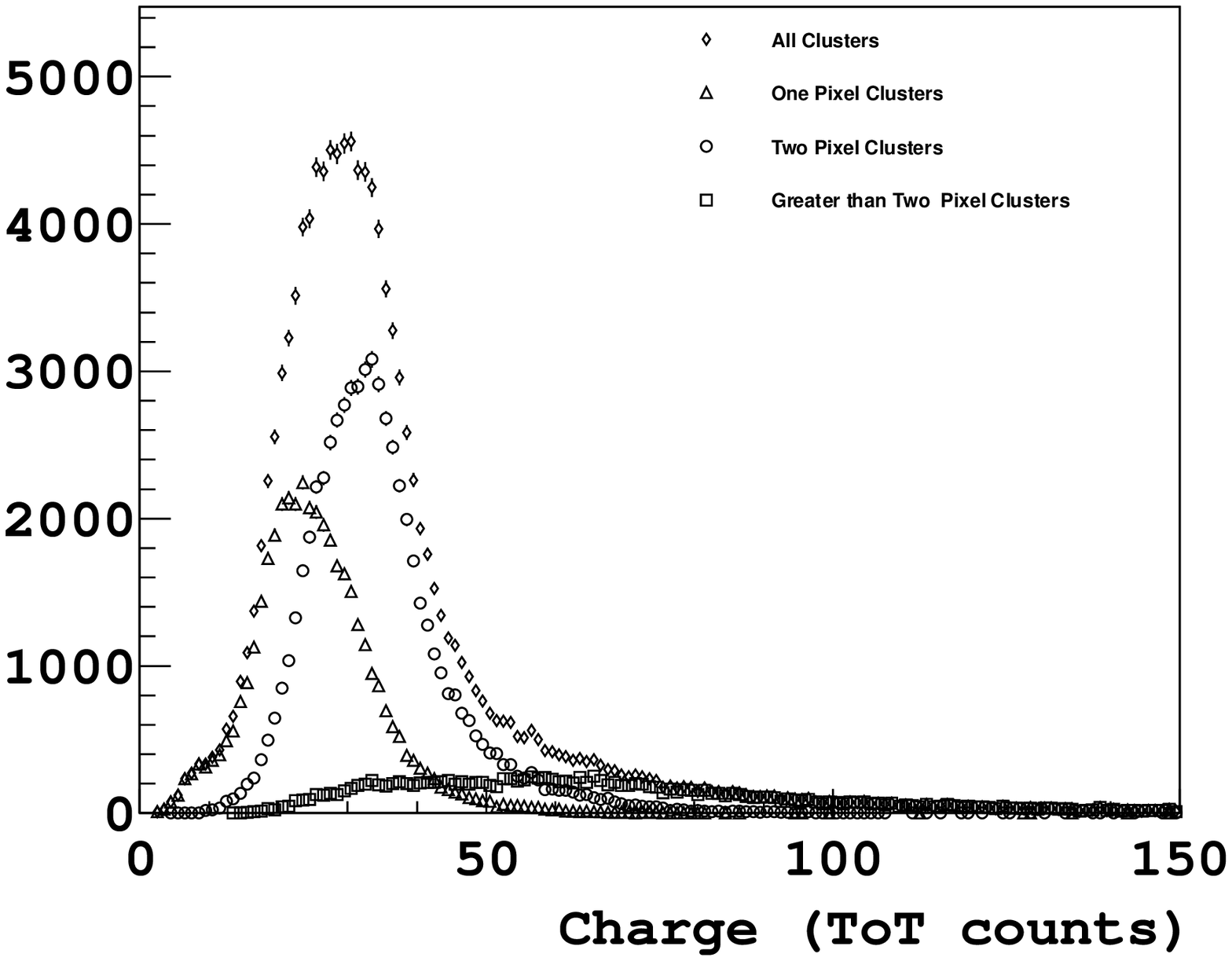}
\caption{Raw ToT distributions in the 3D double sided sensor for perpendicular tracks (left) and tracks with 10 degrees incident angle (right).}
\end{figure}

 The Landau distributions in the 3D sensor show a more complex shape due to the geometry of the
structures within the sensor. The raw ToT distributions are illustrated in Figure~\ref{fig:3dlandau} for
tracks at perpendicular and angled incidence. For tracks at perpendicular incidence there is a
clear double peak structure, originating from tracks which pass between the columns and deposit
charge in the full silicon thickness, and tracks which pass through the length of a column and
deposit charge in a small fraction of the thickness below.  For tracks passing through the sensor
at larger angles of incidence, the distribution represents a more uniform Landau shape. 
In both cases the distribution  is seen to be separated from the noise floor,
but closer to the threshold than in the case of the planar sensor. The detailed
characteristics of the Landau distribution are discussed more extensively in the companion paper~\cite{3Dcompanion}.
\section{Efficiency and noise as a function of threshold}
\label{s:hamish}

To measure the efficiency of the DUT, tracks were reconstructed using all six telescope planes.
The intersection point of each track with the plane of the DUT was calculated, and clusters within
0.2~mm of this point were assumed to result from that track. However, if there was another
track within 0.6~mm of the track in question, both tracks were ignored to avoid associating clusters to the wrong track.
The range of $0.2$~mm is at least 10 times the width of most
residual distributions, chosen so it is highly unlikely that genuine clusters were omitted. 

\subsection{Planar sensor efficiency and noise results}
The efficiency was calculated as a function of threshold for three different angles of the DUT, and is
shown in Figure~\ref{fg:EffAsFcnTHL_301110}. For most thresholds the efficiency is measured to be $99.5\%$, with a
significant drop at higher thresholds. 
At low threshold levels analogue noise can cause individual pixels to be momentarily insensitive to external
signals as the digital part of the pixel only responds to positive transitions across the threshold during the
period when the shutter is open. Thus a high noise rate will increase the chance of the discriminator level already
being above threshold when the shutter is opened, causing the overall efficiency of the device to fall.

The possibility of associating clusters to tracks incorrectly,
and thereby overestimating the efficiency, was considered. The probability of associating a random cluster to a track
where no real cluster existed was evaluated. It dominates the uncertainty in Figure~\ref{fg:EffAsFcnTHL_301110} for
low thresholds, but falls rapidly and becomes negligible at thresholds above 1000~electrons. The plot in
Figure~\ref{fg:EffAsFcnTHL_301110} has statistical uncertainties and this systematic uncertainty added in quadrature.

The noise was also evaluated as a function of threshold for runs taken without beam, and is plotted in
Figure~\ref{fg:NoisevsTHL_041110}. It is high for very low thresholds, as expected, then falls to zero, or almost zero, for most thresholds under test.
\twoepspicx{EffAsFcnTHL_301110}{Efficiency (fraction of tracks
  with associated clusters) of the $55 \times 55$~$\mu$m$^2$ planar DUT as a function of threshold.}
{NoisevsTHL_041110}{Noise (mean number of hits per event in runs without beam) as a function of threshold.}

\subsection{3D sensor efficiency results}
As described in Section~\ref{s:3dsensor}, the Timepix chip used for
the manufacture of the the 3D sensor was of a lower grade and had two
non-responding columns (a total of 512 dead pixels). Furthermore, the bump bonding of the sensor was not perfect
and introduced further dead or noisy pixels. For this reason,
extra cuts were applied in the 3D sensor analysis in order to remove
dead pixels.  A map of dead and
noisy pixels was produced exposing the sensor for 20 minutes to an X-ray source.
An average of 1000 counts was obtained per pixel, and those pixels
more than four standard deviations from the mean were flagged as dead
or noisy.
This identified a total of 640 pixels (including the two
dead columns). This map was used in the analysis and all extrapolated tracks within
three pixels of a dead or noisy pixel on the 3D sensor were excluded from the analysis. 
Furthermore, all extrapolated tracks were required to be within the active
area of the 3D sensor; the extrapolated track position was required to
be seven columns from the edge of the sensor.

The double sided 3D sensor was biased to 20~V. This voltage
ensures that both the inter-column region and the regions beneath the
columns and the front and back-planes of the sensor are biased~\cite{firstDoubleSided3D}. The Timepix chip threshold was set to
approximately 1000 electrons.

\begin{figure}[t]
 \begin{center}
   \includegraphics[width=0.5\textwidth]{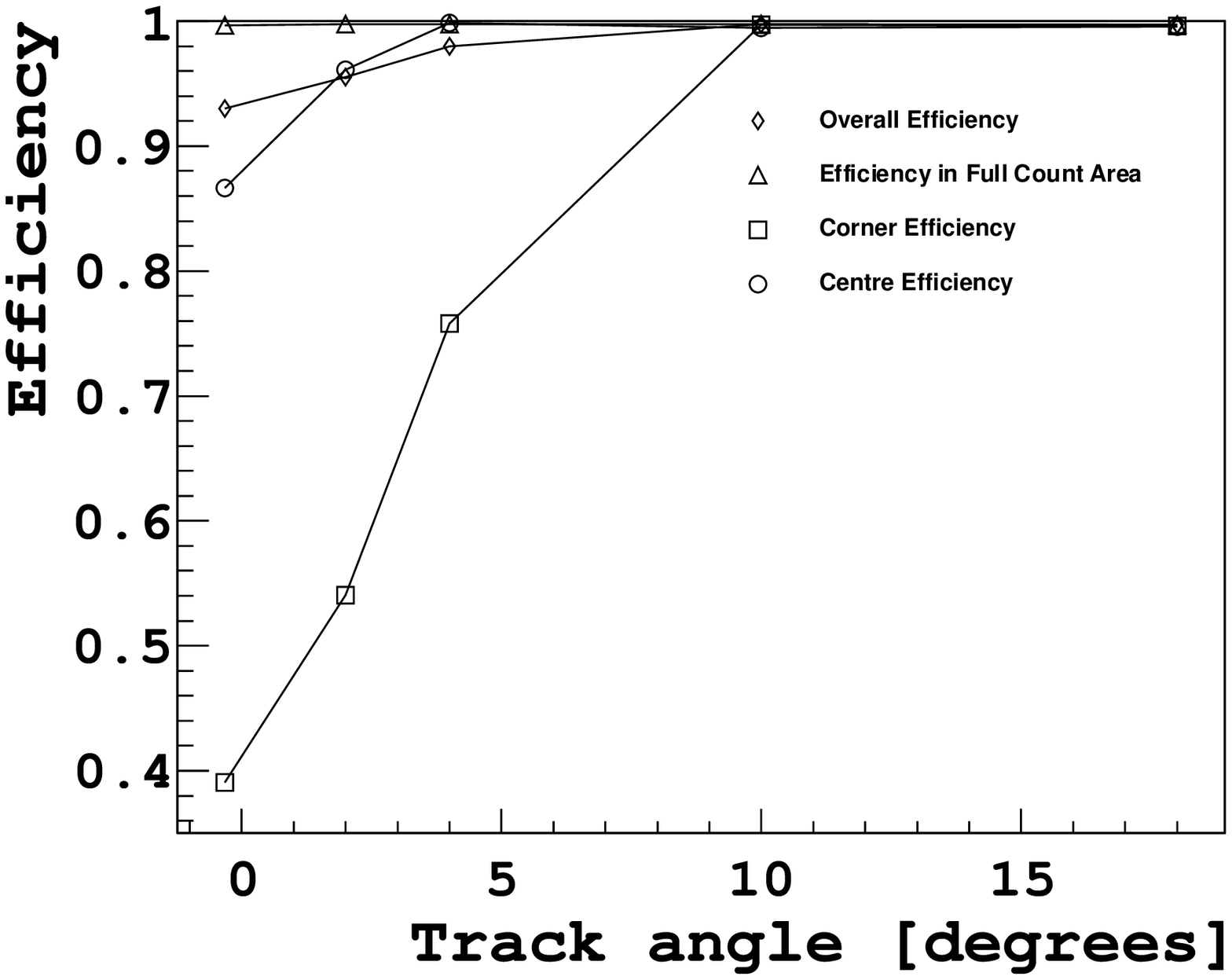}
 \end{center}
\caption{Efficiency of a double sided 3D pixel sensor rotated
 around the \textbf{y} axis. The overall efficiency is shown alongside that
 for three different regions of each pixel cell: the regions around
 the corner and centre electrodes, and the region away from the
 electrodes.}
\label{fg:3DEfficiency}
\end{figure}

Figure~\ref{fg:3DEfficiency} shows the efficiency of the 3D sensor
as a function of the rotation angle of the sensor. 
The uncertainties on the efficiencies 
were estimated from the mis-identification rate.  At a rotation
angle of $10^\circ$ the sensor is fully efficient, with an overall
efficiency measured as $99.8 \%$. 

As the efficiency variation across a unit pixel cell of the 3D sensor
at perpendicular incidence is the main topic of the companion
paper~\cite{3Dcompanion}, only a brief description is provided here. 
As discussed in the introduction, the 3D sensor has a 
series of inactive columns, and for a perpendicular beam no charge would
be collected in the centre of these columns. However, the columns in
the double sided 3D sensor have $50$~$\mu$m of silicon above them
which will collect charge of around 4000 electrons when the sensor is biased at more than
a few volts. As the threshold of the sensor in this study is 1000 electrons,
the columns in the centre of the sensor are expected to be
reasonably efficient, as seen in Figure~\ref{fg:3DEfficiency}, where
the efficiency in a region of $6.25$~$\mu$m radius around the centre columns
is shown. However, for the columns at the corner of a pixel cell, the
charge is shared between neighbouring pixels and the efficiency is
lower, as seen in Figure~\ref{fg:3DEfficiency} for a selection of the
same size region around the corner electrodes. As expected the
remaining region of the cell away from the corners or centre 
remains highly efficient.  The sensor is aligned using the weighted
time over threshold information in the cluster positions, which leads to the pixel cell
positions being defined at the mid-plane of the silicon.  
The non-uniformity effect for the corners and centres is removed as the sensor is rotated, as
seen in Figure~\ref{fg:3DEfficiency}. For this reason it is expected that the 3D sensors
will be tilted when used in a barrel detector, whereas a forward
detector geometry leads naturally to an average track angle.

\section{Angle dependence}
\label{s:anglescans}
\subsection{Introduction}
\label{s:angleintro}
As described in Section~\ref{s:analysisofdeviceundertest} the incident angles of the tracks were
varied in two directions: the ``normal'' direction, in which the chip was mounted perpendicular to the
floor and rotated about the global \textbf{y} axis, and the ``diagonal'' direction, where it was first rotated by 45$^\circ$
about the \textbf{z} axis and subsequently rotated about the global \textbf{y} axis. 
The coordinate system is shown in Figure~\ref{fg:marco} and the measurements performed for the planar
and 3D sensor are described in the following sections. 
\epspicz{1}{2}{marco}{Coordinate system for rotation in the ``normal'' direction (left figure) and the ``diagonal'' direction (right figure).}
\subsection{Corrections due to non-linear charge sharing}
\label{s:etacorrections}
Due to the fact that the pixel pitch is large compared to the diffusion width of drifting holes in 300~$\mu$m silicon,
the charge sharing between cells is not perfect and the simple weighted charge described in Section~\ref{s:clustering} is not
expected to reproduce accurately the track position. This effect is a strong function of angle. In order to study this effect
the precise pointing resolution of the tracks was used to scan the cluster characteristics across the pixel cells, and the
results are shown in Figures~\ref{fg:clustercontributions}~and~\ref{fg:etadistribution} for a selection of runs where the
DUT was oriented at $0^{\circ}$, $5^{\circ}$, $8^{\circ}$ and $18^{\circ}$ relative to the perpendicular. In Figure~\ref{fg:clustercontributions} the track
position is used to determine the inter-pixel distance, and for each position the contribution of one, two, and three column pixel clusters
is displayed. For the perpendicular tracks there is a contribution of two column clusters from the area centered between the pixels,
and close to the pixel centres only one column pixels contribute. As the angle is increased, the charge sharing region increases,
until an optimum close to $9^{\circ}$. After this point the contribution of three column clusters increases. 
By comparing the weighted charge position to the track position the effect can be corrected in the data, as displayed in
Figure~\ref{fg:etadistribution}. Each cluster is used to define the relevant pixel cell, and for this pixel cell the track
position is plotted relative to the weighted charge position. The resulting histogram is fitted with a fifth order polynomial,
and the result is used to correct the weighted charge position in the data.
For angles above $12^{\circ}$ a separate correction is applied to the two column and greater than two column clusters.
This correction, referred to in the rest of this paper as ``eta
correction'',  results in an improved resolution, as discussed in Section~\ref{s:resolution}.
\epspicz{1}{1}{clustercontributions}{Contribution of one, two, and three column width pixel clusters in the dut,
for a selection of runs with the DUT oriented at $0^{\circ}$, $5^{\circ}$, $8^{\circ}$ and $18^{\circ}$.}
\epspicz{1}{1}{etadistribution}{Eta corrections to the data. The comparison of the track position to the charge weighted cluster centre
is shown for four sample angles, before and after correction. A separate correction is applied to two and three column width clusters for the 18$^\circ$ plots.}
\subsection{Planar sensor normal angle scans}
\label{s:resolution}

Figure~\ref{fg:resbeforeaft} shows the residual distributions
as a function of incident track angle
together with the contributions of the one, two, and three column pixel
clusters, before and after eta correction.
These data are referred to as ``Normal Angle Scan'' as the track
incident angle is varied across the pixel columns, as opposed to the
diagonal angle scan described in the next section.
It can be seen that for the perpendicular tracks there is a good improvement in the double column clusters
after the eta correction. As the track angle increases the improvement gets smaller, giving negligible improvement
close to the optimal 9 degree angle. For large angles there are two main categories of cluster, two and three column clusters.
The three column clusters give fairly good resolution in the regions closer to the pixels, where the spatial information
is dominated by two external pixel hits. For the tracks centered between the pixels, the clusters are predominantly
two column clusters.  The resolution here is slightly worse, as the two columns tend to have a similar pulse height.
For these larger angles the eta correction gives only a modest
improvement.

\begin{figure}[p]
\centering
\includegraphics[width=0.8\linewidth]{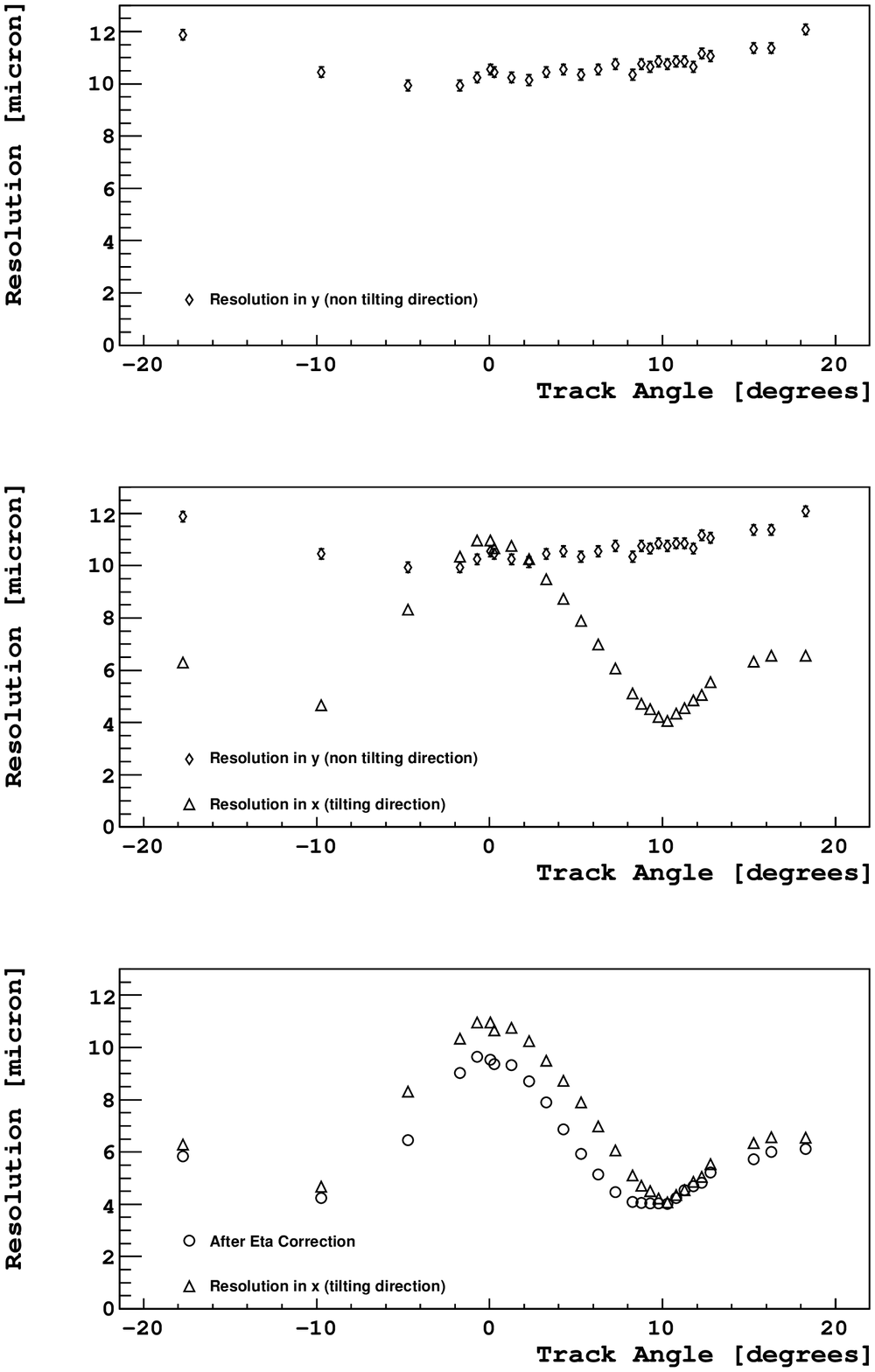}
\caption{Resolution of the $300~{\rm \mu m}$
  thick planar sensor 
$55~{\rm \mu m} \times 55~{\rm \mu m}^2$ pixel DUT 
as a function
  of track angle. The  2.3~$\mu$m  track
pointing resolution has been subtracted. From top to bottom : in the non-tilting direction, in the non-tilting
compared to in the tilting direction, before and after eta correction in the tilting direction.}
\label{fg:tiltnotilt}
\end{figure}

The resolution of the DUT is extracted from the single Gaussian fit,
and the $2.3 {\rm \mu m}$ contribution of the track
(as discussed in Section~\ref{s:pointingresolution}) is subtracted.
The angle of the track is corrected by $0.29^\circ$ according to the analysis described in
Section~\ref{s:perpendicularpoint}.  The raw resolution of the planar sensor DUT is shown for the tilting and
non-tilting direction in Figure~\ref{fg:tiltnotilt}.  In the
tilting direction a dramatic improvement is seen as a function of
track angle, with a best resolution of 4.0~$\mu$m for tracks oriented at 8-$10^{\circ}$ and a worst resolution of $\approx$11~$\mu$m.
In the non-tilting direction there is a small
degradation as the angle increases, due to the fact that the information is shared over more
pixels, which may fall below threshold.

The improvement brought by
the eta correction described in Section~\ref{s:etacorrections} is also illustrated in
Figure~\ref{fg:tiltnotilt}.
The best improvement is achieved when the tracks impinge perpendicularly on the
sensor; for the optimum resolution the eta correction does not bring
an improvement.
For the data shown here the full individual pixel gain calibration
described in Section~\ref{s:calibration} has
applied, however note that in terms of the best acheivable
resolution this calibration has only a small impact;
%
after eta correction the individual pixel gain calibrated data have a
better resolution by approximately 0.2~$\mu$m compared to the
uncalibrated eta corrected data.
The reason for
this lies in the shape of the eta distribution itself: the offset
introduced by the Timepix partially compensates the intrinsic skew in
the eta distribution from the silicon response itself.  When this
offset is removed by applying the calibration the resolution worsens.  After eta correction, which
absorbs most of the effect, the resolution in both cases is very comparable.

\epspicz{1}{1}{resbeforeaft}{Examples of the residual fitting to the DUT before and after eta correction, for four sample orientations of
the DUT ($0^{\circ}$, $5^{\circ}$, $8^{\circ}$ and $18^{\circ}$ ).
The residuals of all clusters are shown, and the residuals of one, two, and three column clusters are shown separately.}

\subsection{Planar sensor diagonal angle scans}
\begin{figure}[htb]
\centering
\includegraphics[width=0.45\linewidth]{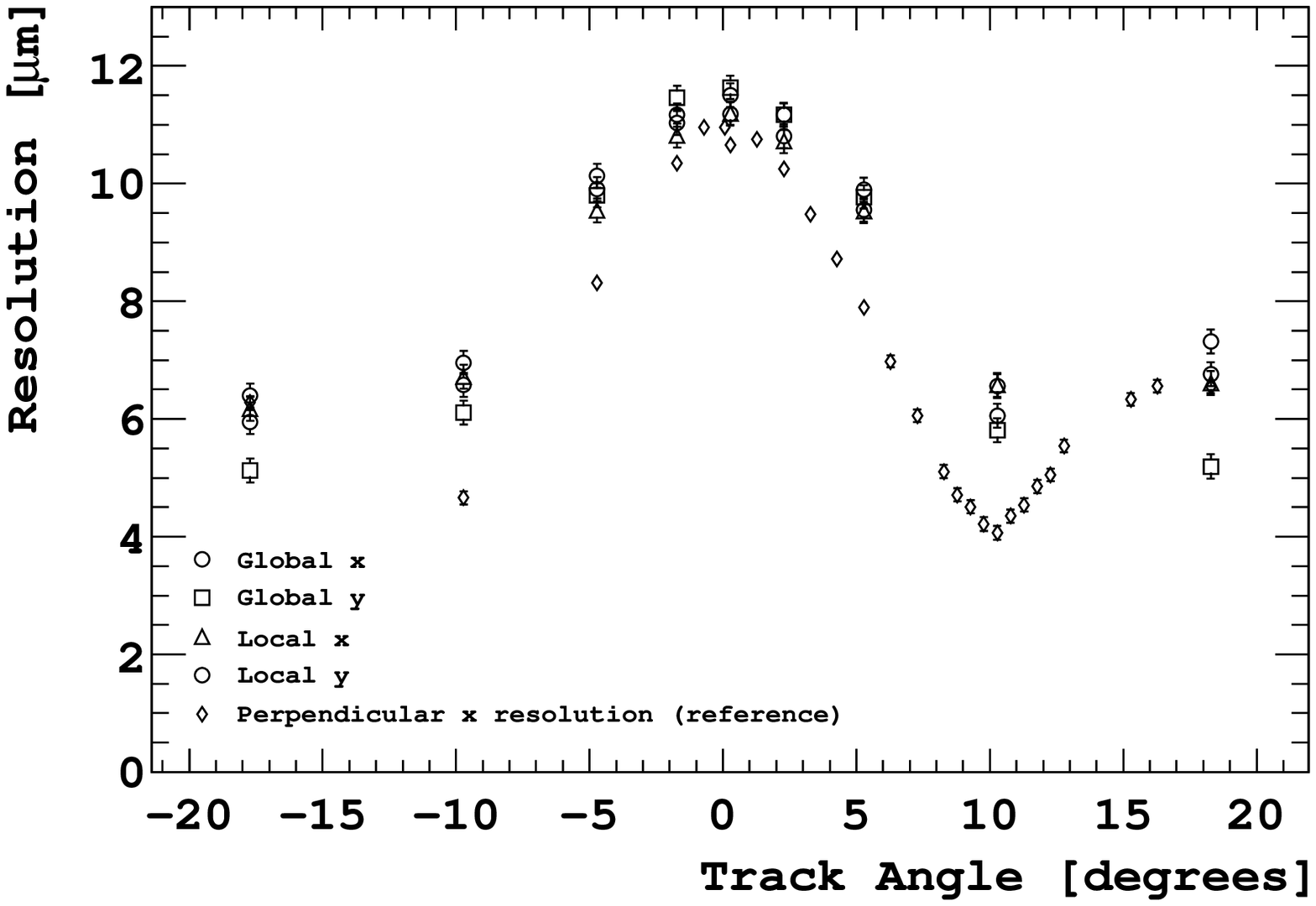}
\includegraphics[width=0.45\linewidth]{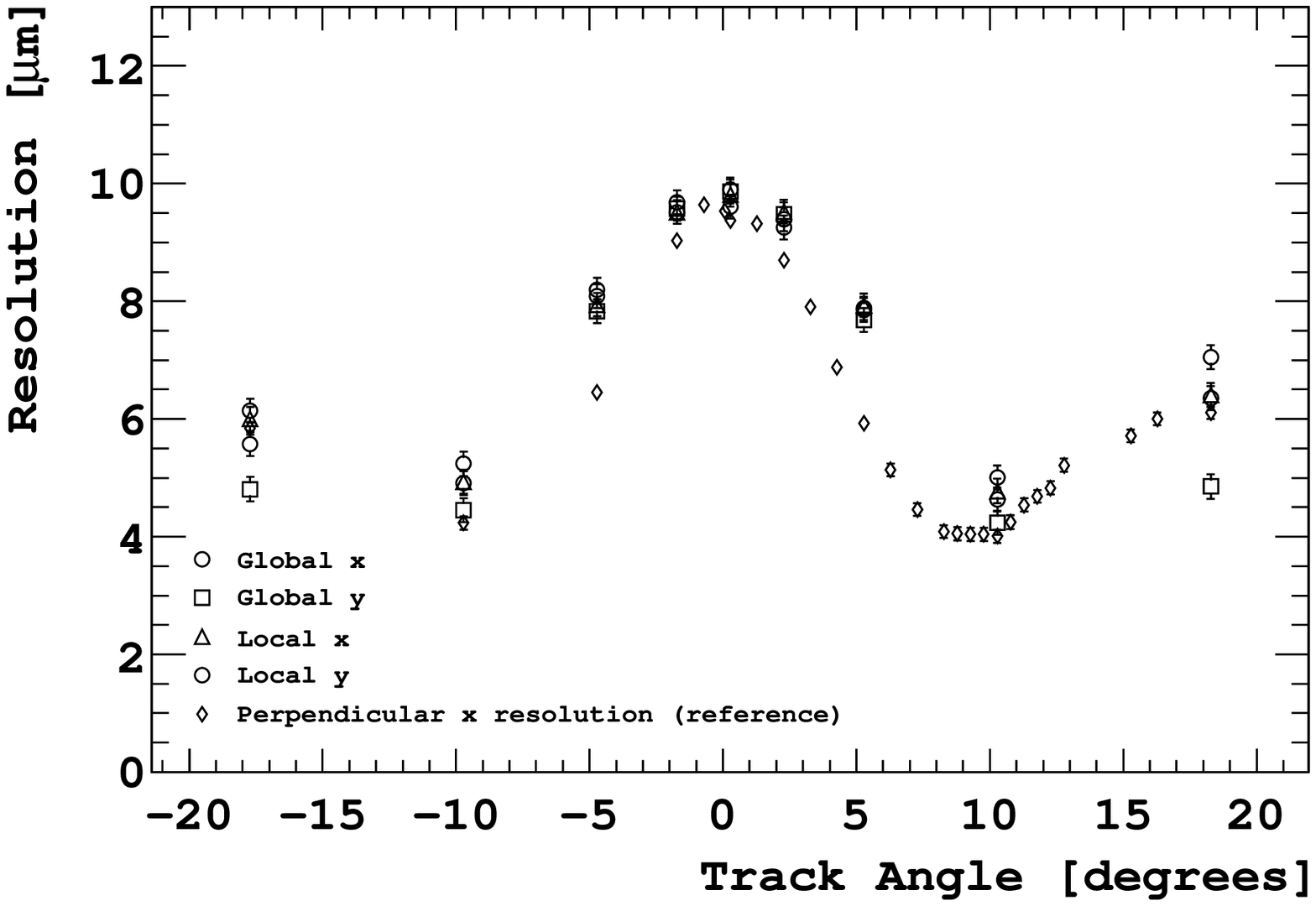}
\caption{The resolution in the diagonal and normal directions as a function of
track angle variation in the diagonal direction (left), and the resolution after the eta correction
is applied (right). The resolution improvement is a strong function of the
angle, giving a best improvement of around 2${\rm \mu m}$ close to perpendicular.}
\label{fg:diagonaleta}
\end{figure}

Given an equivalent ``normal'' resolution of the pixel device in the \textbf{row} and \textbf{column} direction,
it is expected that the ``diagonal'' resolution, i.e. between pixels displaced by one in column and row,
should also be equivalent.  However, in this direction the shapes of the pixel clusters
are different, which could lead to differences in the results given by the charge weighted centre method, and an
eta correction is required in both the \textbf{x} and \textbf{y} direction, which could also affect the result. In addition, the
effective pixel pitch in this direction is greater by $\sqrt(2)$ which can lead to a different rotation angle
for the minimum resolution.  This was explicitly checked in the
system by rotating the DUT by $45^{\circ}$ around the \textbf{z} axis,
such that the ``x'' (\textbf{column}) and ``y'' (\textbf{row})
measurements in the telescope were probing the ``u'' and ``v'' directions in the DUT, rotated by $45^{\circ}$
relative to the \textbf{row} and \textbf{column}
directions. For each run, the raw resolution after pixel correction is plotted,
and an eta correction is determined in the same way as in Section~\ref{s:etacorrections} for both the
\textbf{row} and \textbf{column} direction. 
The resolution is then redetermined after applying the eta correction. Each measurement is based on approximately
100,000 residual measurements in the DUT. The results are shown in Figure~\ref{fg:diagonaleta}.

The left plot shows
the raw results, with no eta correction applied. The resolution is shown as a function of the changing track angle
in the diagonal direction.  For reference the equivalent result obtained for the ``normal'' resolution measured
as a function of track angle in the normal direction, is shown.  For perpendicular tracks the diagonal resolution
is equivalent to the ``normal'' resolution, although the shape of the residual histogram (not shown here) is
different, due to the contribution of different cluster shapes.  As the track angle increases in the diagonal
direction, there is an improvement seen for normal tracks both in the direction of rotation, but also in the
perpendicular direction, which might not be naively expected.  This can be understood by considering the pixel
layout: as the sensor rotates information is picked up from the next, staggered row of pixels, so new information
contributes in both coordinates.  

Improvements in resolution as a function of the track diagonal angle are
also seen for the ``normal'' resolutions.  It can be seen that the shapes of the curves differ, and in the
diagonal direction the resolutions are not as good as that achieved for the best ``normal'' resolutions. 
The right plot shows the same curves after eta corrections
are applied in the local \textbf{row} and \textbf{column} directions. It can be seen that for the diagonal data the eta
corrections give similar improvements to the normal data. 
\subsection{3D sensor normal angle scans}
The resolution of the double-sided 3D sensor is shown in
Figure~\ref{fg:3DResolution} as a function of the angle of rotation of
the sensor. As shown in~\cite{3Dcompanion}, the 3D sensor has
relatively little charge sharing compared with a planar sensor. This
is expected from the self-shielding electric-field geometry of the 3D
sensor. In a planar sensor charge drifts through the thickness of the
sensor to the collection electrodes and diffusion naturally leads to
charge sharing. In a 3D sensor the charge drifts to the collection
electrode column and hence tends to remain bounded inside a pixel.
As the angle between the sensor and the tracks is varied, the
geometry leads to the charge being deposited across multiple
pixels. The fraction of single and multiple pixel clusters is shown in
Figure~\ref{fg:3DFractionClusters}. At perpendicular incidence,  the measured 3D Sensor
resolution is $15.8$~$\mu$m, which is compatible with the expected binary
resolution for a $55$~$\mu$m pixel. The optimal resolution obtained was
at a rotation angle of $10^\circ$, where a  resolution of $9.2$~$\mu$m
was measured. This angle corresponds to the entry and exit points of
the track in the silicon being separated by approximately one pixel,
and hence maximises the two pixel fraction.  No eta correction has
been applied in this analysis, due to the limited amount of diffusion seen.
\twoepspicx{3DResolution}{Resolution of a the double sided 3D
  pixel DUT as a function of track incidence angle.}
{3DFractionClusters}{Fraction of single and multiple pixel
  clusters in the double sided 3D pixel DUT as a function of the rotation angle.}

\section{Planar sensor bias voltage scans}
\label{s:HVResolution}

The effects of bias voltage on the DUT performance were investigated by taking a
number of measurements with the voltages ranging from 5 to 200V.  Each 
scan was performed by ramping the HV both up and down, to isolate
affects solely associated with the bias conditions of the sensor.  As
described in Section~\ref{planarsensor} the nominal depletion voltage
of the device was 10V.  This was confirmed by exposing
the device backside to $\alpha$ source and measuring the voltage at which the signal from the energy deposited by
the $\alpha$ particle reaches the collecting electrodes.



\begin{figure}[htb]
\centering
\includegraphics[width=0.7\linewidth]{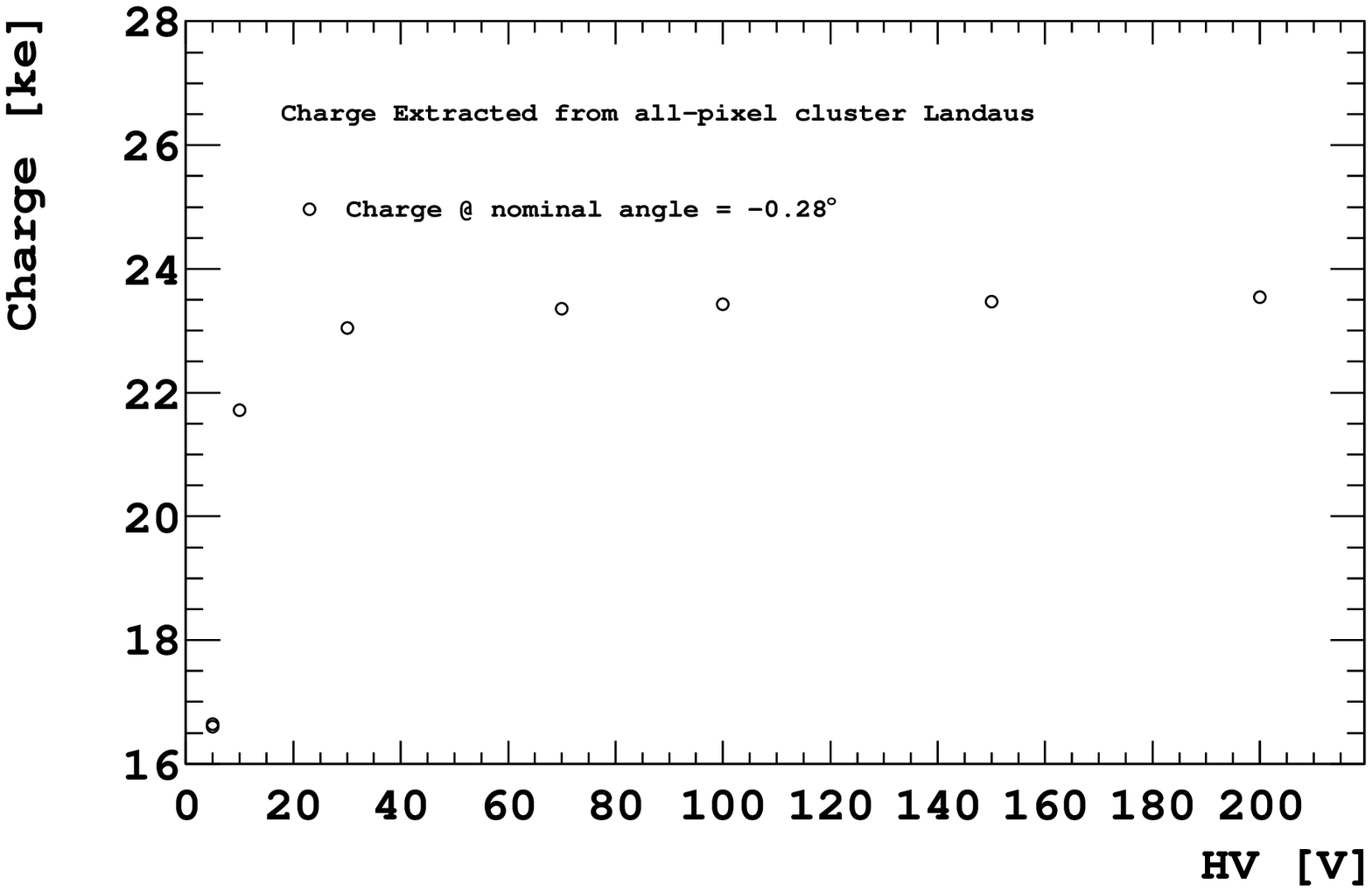}
\caption{ Most probable value of the collected charge for tracks at normal incidence as a function of HV.}
\label{fig:Landau}
\end{figure}

Figure~\ref{fig:Landau} shows the evolution of the measured most probable value of the cluster charge for tracks at normal
incidence. The ToT to charge conversions were performed as described
in the same way as described in Section~\ref{planarsensorlandau}, and
the most probable values are determined with fits to Landau curves
convoluted with a Gaussian described before.    Above the depletion
voltage the curve rapidly reaches the maximum value of a
about 23,500 e$^-$.  This is expected, as the peaking time of the
electronics (around 100~ns) is much bigger than the typical collection
time in 300~$\mu$m planar sensors.

\begin{figure}[htb]
\centering
\includegraphics[width=0.7\linewidth]{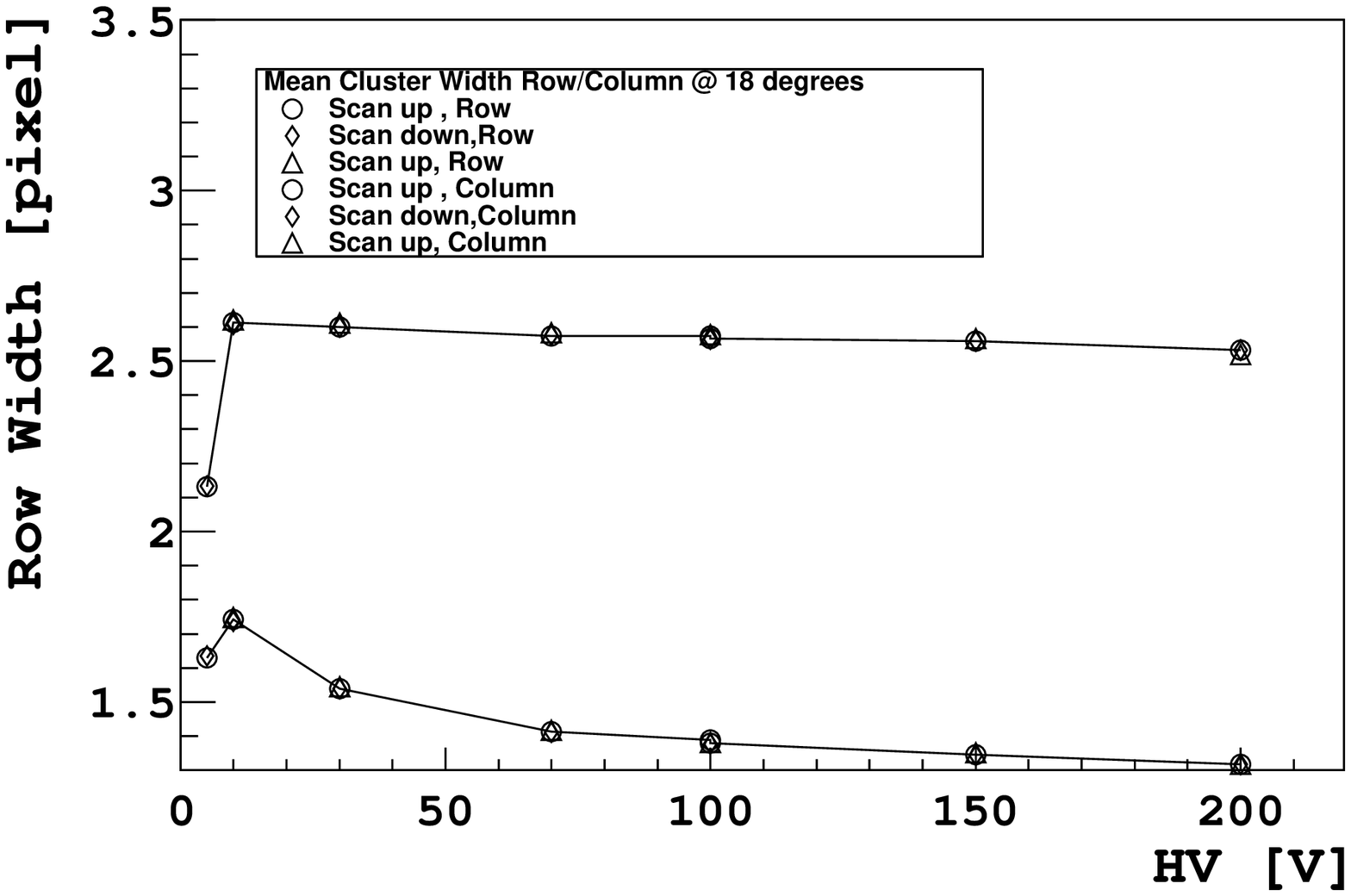}
\caption{  Cluster width along the tilted (row) and non-tilted (column) as a function of HV. Note that the maximum cluster
size is achieved when the sensor reaches full depletion. In the non-tilted direction, the cluster size decreases
significantly as the sensor becomes more and more overdepleted.}
\label{fig:clustersize}
\end{figure}

Figure \ref{fig:clustersize} shows the evolution of the cluster size as a function of the HV, in the bending plane
(column projection) and non-bending plane (row projection). In the projection where the
particle is at normal incidence, the cluster size is 
maximum when the depletion voltage is reached, and then it decreases as the bias voltage increases because of the
faster collection time. Eventually the charge drift speed reaches saturation and the cluster size reaches an asymptotic value.
In the projection where the track is inclined, and thus the cluster
size is dominated by the geomtrical factor, there is no change in the
cluster width above the depletion voltage.

\begin{figure}[p]
\centering
\includegraphics[width=1.0\linewidth]{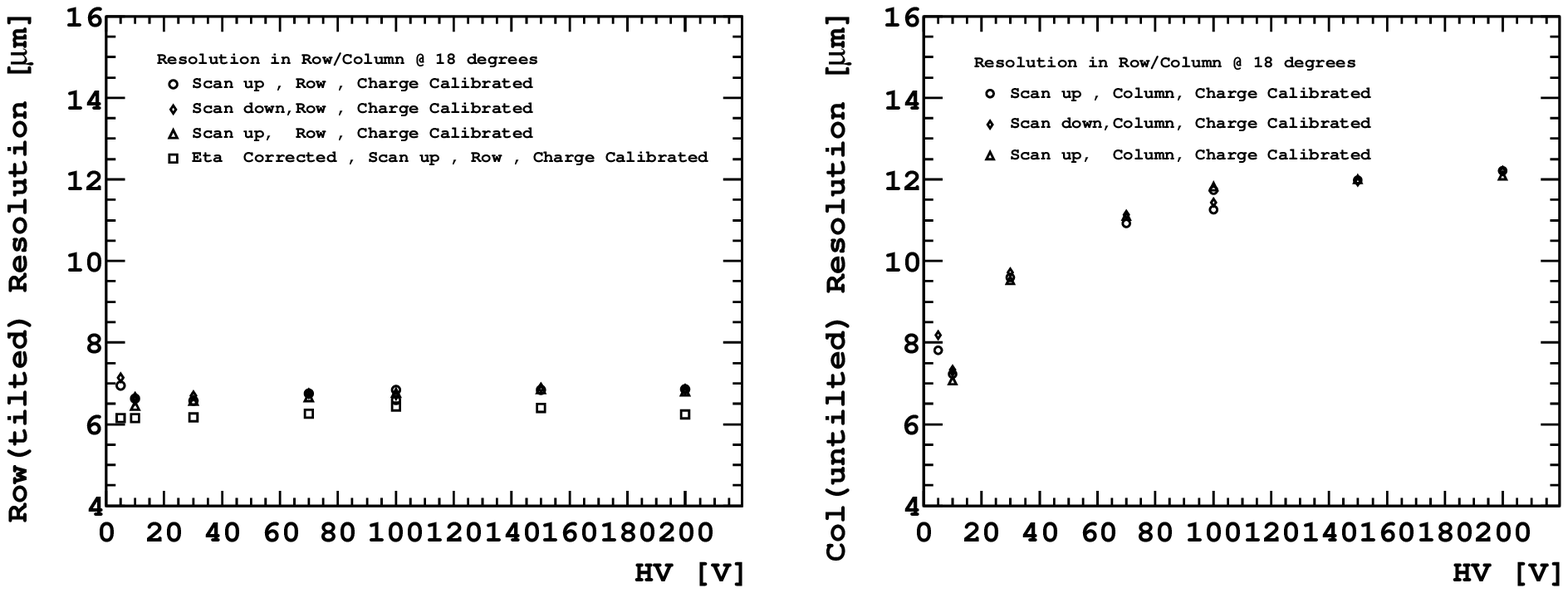}
\caption{ Spatial resolution as a function of the applied bias voltage with the DUT oriented at 10$^\circ $ with respect
to the beam axis along the column direction. On the left  the resolution for the coordinate which is sensitive to the
sensor angle is shown before (circles and diamonds), and after
(triangles and squares) eta corrections are applied; on the right
figure the corresponding curves along the non-tilted directions are
shown.   The effect of the larger diffusion radius in improving the
charge sharing at low bias voltage is clearly visible for the
perpendicular incidence tracks; for the tracks impinging at 10$^\circ$
the geometrical factor dominates and the resolution is independent of
applied voltage.}
\label{fig:reso}
\end{figure}

\begin{figure}[p]
\centering
\includegraphics[width=0.7\linewidth]{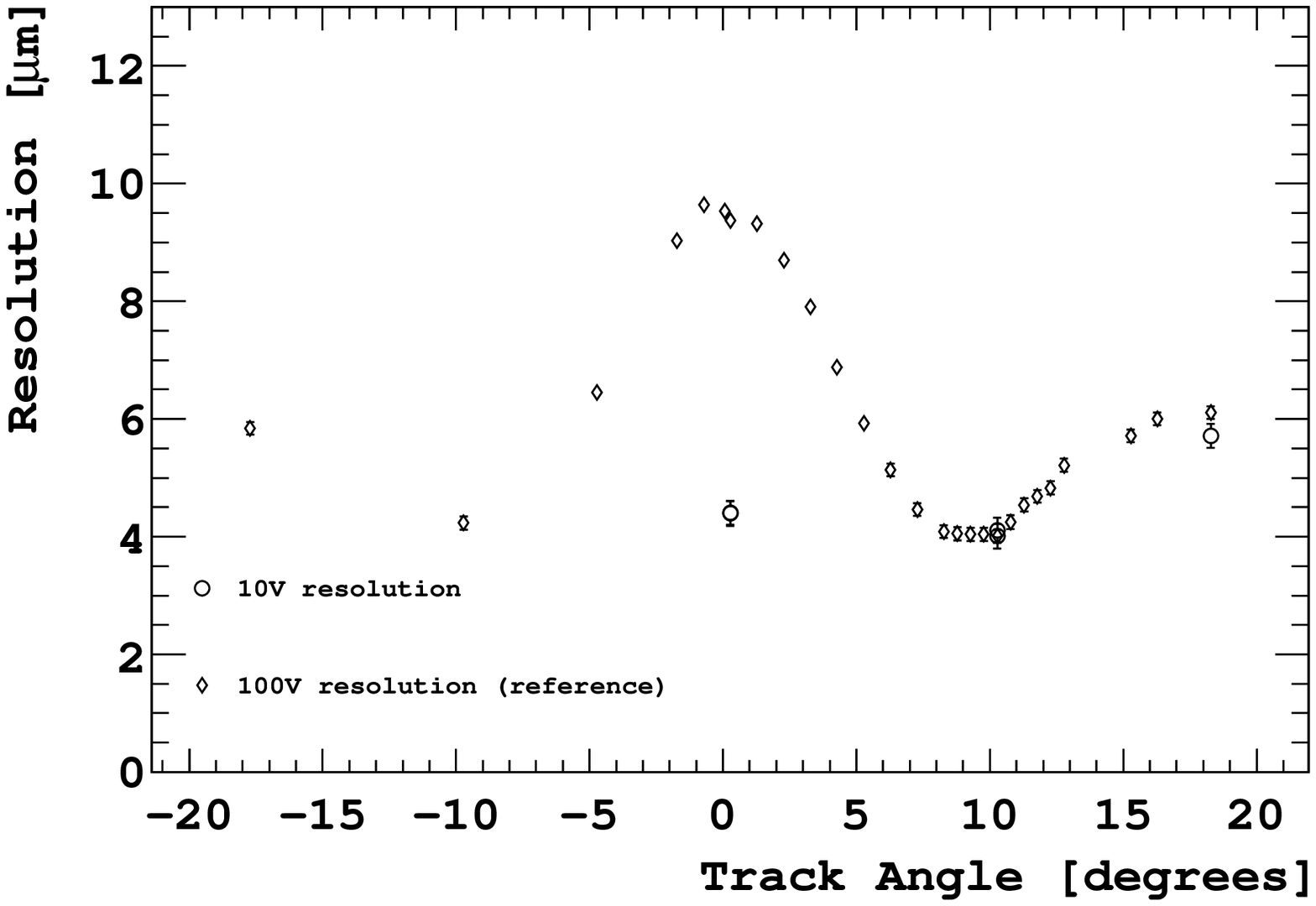}
\caption{Comparison between the spatial resolution achieved with the sensor biased at depletion voltage
(10~V) and overdepleted (100~V), as a function of track incident angle.}
\label{fig:anglecomparison}
\end{figure}

Figure \ref{fig:reso} shows the corresponding trends in spatial
resolution, extracted in the same way as described in
Section~\ref{extractingresolution}.
%
%
Note that
in the angled projection, the resolution reaches a maximum when the sensor is fully depleted, and then it remains
essentially unchanged. Conversely, in the projection at almost normal incidence, the additional charge diffusion made
possible by the slower collection time, provides a significant improvement in the resolution.  
This effect is illustrated in Fig. \ref{fig:anglecomparison} which compares the resolution as a function of angle when
the DUT is biased at 10~V  with  the angle scan results presented above,  which was  taken with sensor over-depleted
(V$_{bias}=100$~V). The improvement at low angles from the greater charge spread due to diffusion is evident.
\section{Time-walk}
\label{sec:timewalk}
During the testbeam several runs were taken with the DUT in Time-of-Arrival (ToA) mode. 
These runs were taken to determine the spread in arrival time of pixels within the same cluster.
In view of the VeloPix developments the spread in the ToA should be smaller than one LHC bunch crossing
time (25~ns), as this will simplify the collection of the event fragment belonging to the same bunch crossing\footnote{The
measured time of arrival is highly correlated to the time over threshold. Hence correction of ToA using ToT information is
possible but unwanted because it adds complexity to the processing of data in the off-detector electronics.}.

\begin{figure}[p]
\begin{center}
\includegraphics[width=6in]{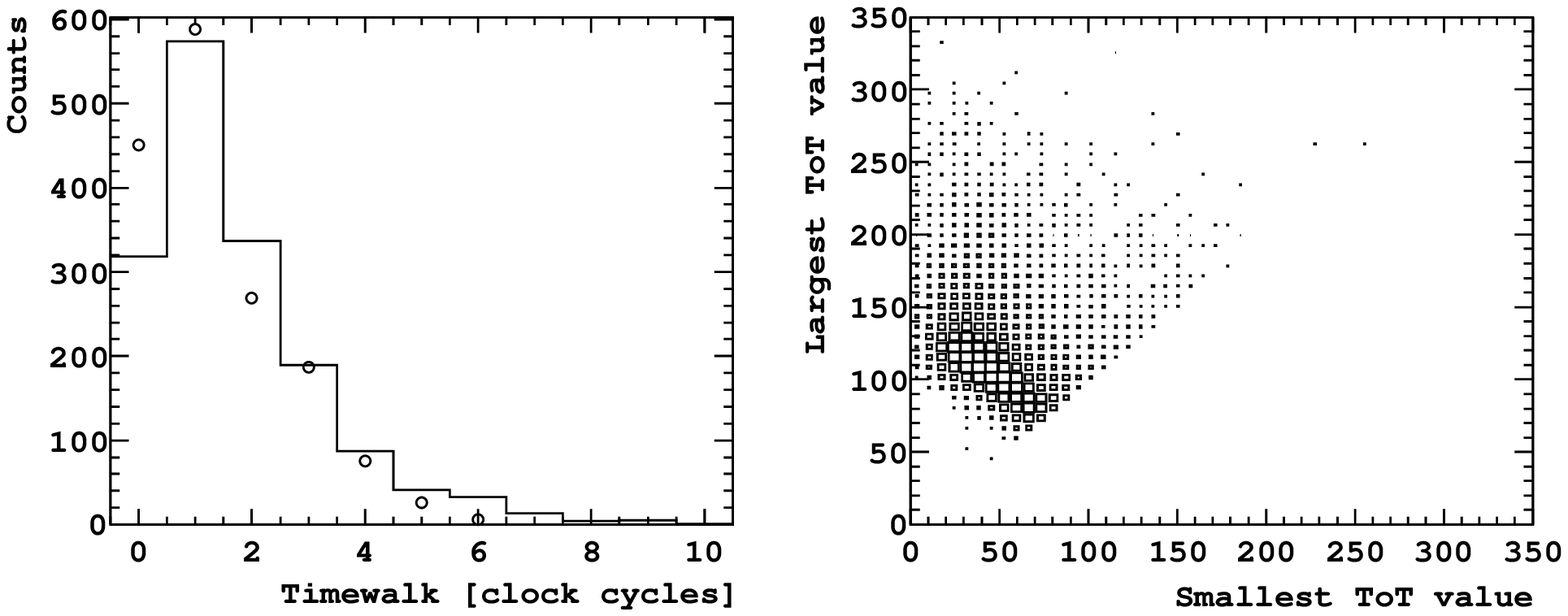}
\caption{The left plot shows the measured difference in ToA the hits in a 2-pixel cluster (histogram)
and the calculated ToA difference (open circles) using the ToT distribution shown in the right hand plot.}
\label{fig:tw1}
\end{center}
\end{figure}

A source of this time spread is the timewalk of the discriminator which is largely due to the non-zero rise-time of the pre-amplifier output signal.
The maximum timewalk can be defined as the time difference between the earliest possible discriminator output signal (for
large energy depositions) and that for a signal which just crosses threshold. An ideal discriminator will have a maximum
timewalk equal to the rise-time of its input signal. 
However, for the Timepix chip the measured difference in arrival time of hits within the same cluster can be as large
as 275~ns, as shown left plot of Figure~\ref{fig:tw1}.

This large timewalk, which is significantly larger than the specified rise-time of the preamplifier of ~100 ns, is also due to a lack of overdrive of the discriminator input.
For a small voltage difference between signal input and threshold input the high frequency gain of the discriminator is not
sufficient to make the discriminator output swing from power rail to power rail. As a consequence the discriminator output will
only reach its logical switching level at a later time\footnote{The output will swing eventually if the input signal is over
threshold for a long enough time because the gain of the discriminator is much higher at low frequencies.}.

\begin{figure}[p]
\begin{center}
\includegraphics[width=4in]{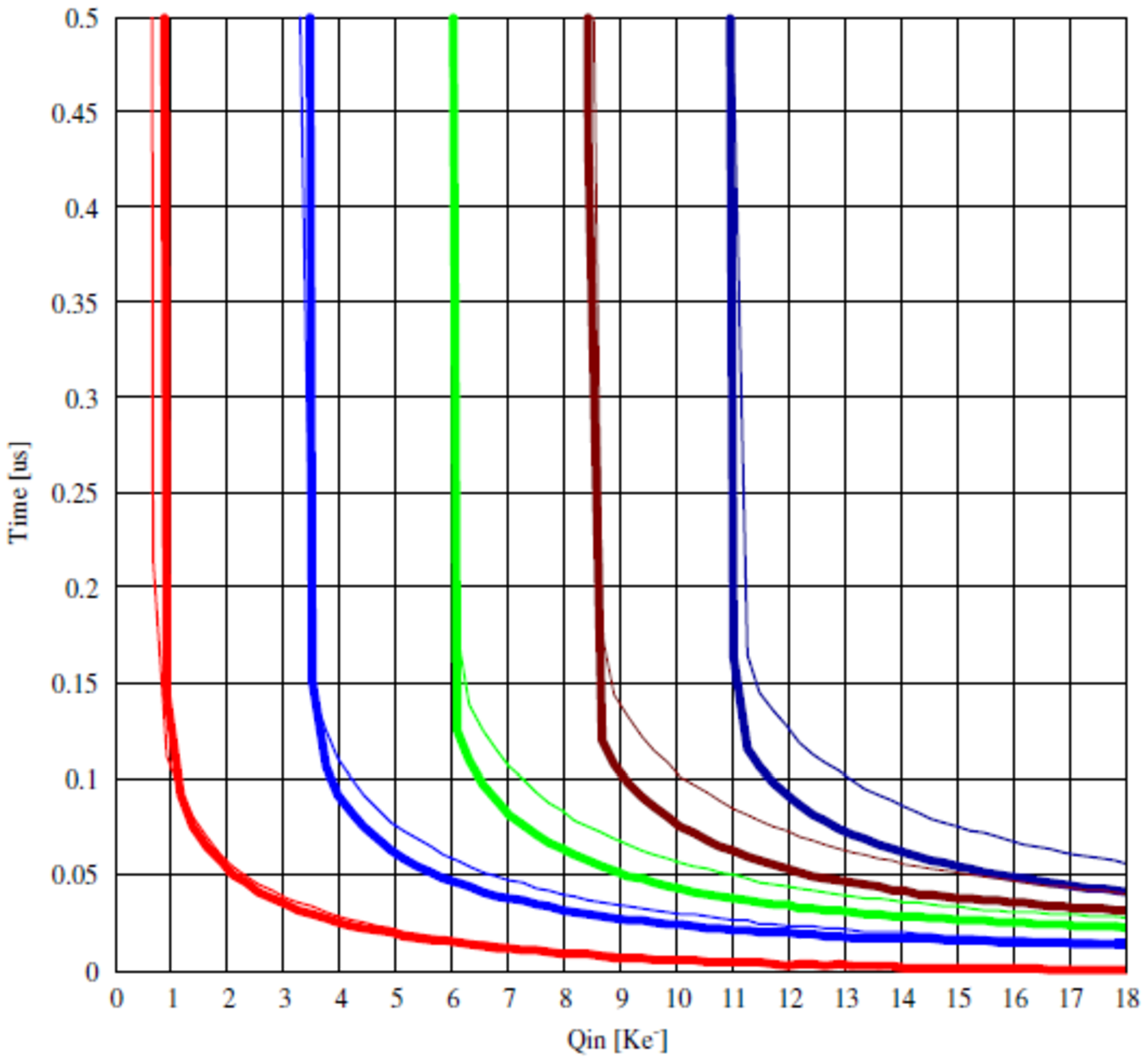}
\caption{Delay time of Timepix front-end as function of input charge for different thresholds (colors), from\cite{timepix}.}
\label{fig:tw2}
\end{center}
\end{figure}

The question is whether the observed difference in ToA is compatible with timewalk of the Timepix chip obtained
from testpulse measurements~\cite{timepix}.
Unfortunately, the Timepix can only measure either the ToA or ToT, not both at the same time. Hence checking the spread
in ToA is done using data from a ToT run in identical running conditions and converting this data using a parametrization of the timewalk from~\cite{timepix}.
\\
The right plot of Figure~\ref{fig:tw1} shows the ToT for the two pixels in a 2-pixel cluster for a run with the threshold set to 400. 
For each bin in this plot, the two ToT values are converted to charge using the surrogate function as described in Section~\ref{s:calibration}.
Next the corresponding time delay of the discriminator is calculated using the charge versus delay curves of the Timepix shown in Figure~\ref{fig:tw2} (from~\cite{timepix}). 
The difference in delay time is modified by adding a 25~ns offset for $50\%$ of the statistics. This offset is needed because the timing clock is inverted from pixel to pixel. 
Hence one of the (neighbouring) pixels in the 2-pixel cluster is delayed by half a clock cycle. 
As the ToA is measured in units of clock cycles, the difference of a half a cycle is rounded to either 0 or 1 depending of arrival time of the particle w.r.t. the phase of the timing clock. 
This clock inversion causes the peak at 1 of the ToA difference distribution.
The open circles in Figure~\ref{fig:tw1} are the result of the predicted timewalk using ToT data. 
The calculated ToA distribution has a shorter tail than the measured distribution. 
However, the tail of the ToA distribution is due to hits with very small energy depositions. For these hits both the uncertainty in ToT and the uncertainty in delay time of the discriminator (and also the error in its parametrization) are largest.
\section{Pulse-shape}
As explained in Section~\ref{s:timepixchip} the response of the
preamplifier can be approximated to a triangular pulse,
with a rise time of the order of 100~ns for the settings used in the 
experimental setup, and a fall time of the order of several
microseconds.  Using the threshold scan data it is possible to
investigate qualitatively the shape of the pulse falling edge by
plotting the raw ToT response as a function of the threshold setting.
For each
threshold setting clusters were selected in the device under test
within 100~$\mu$m of an impinging track.  The ToT for each threshold
setting was determined by fitting a Landau convoluted with a Gaussian,
as described in Section~\ref{s:HVResolution}, and extracting the MPV.
The values were plotted against the threshold setting in electrons,
for all clusters, and for subsamples of one pixel clusters and two
pixel clusters.  The results are shown in Figure~\ref{fg:pulseshape}.
It can be seen that, for the single pixel clusters, there is a
relatively linear behaviour down to thresholds of 3000 electrons or
fewer, after which the non-linear part of the surrogate function is
reflected in a change of shape.  For the double pixel clusters,
at low thresholds and high thresholds the curve lies on opposite sides
of the single pixel curve. The reason for this can be again
understood by considering the offset of the surrogate function.  At
low threshold values there is a positive offset, so two pixel clusters
exhibit a higher total ToT value than single pixel clusters, whereas
at high thresholds the effect is reversed, as the offset decreases and
eventually becomes negative.  For a perfectly working Timepix, two
pixel clusters should always give slightly smaller ToT, corresponding
to twice the threshold value. This plot emphasises the importance of a
good understanding of the surrogate function for the appropriate
threshold in order to extract correct calibrations from Timepix data.
\epspicz{2}{3}{pulseshape}{Timepix pulseshape falling edge, for all clusters, single pixel clusters, and double pixel clusters. The errors on the measured values are negligible.}
\section{The influence of charge digitization on the spatial resolution}
\label{s:gainsection}
For future readout chip designs based on Timepix (such as the proposed VeloPix),
it is important to understand the degradation of the spatial resolution due to limited dynamic
range and charge resolution within the charge digitisation circuit. If the dynamic range spans
less than the full landau spectrum, charge beyond the end of range will accumulate in the highest
digitised count and distort the position reconstruction. A limited charge resolution will also lead
to greater errors in position reconstruction.

Data sets with lower range or resolution were produced by scaling the measured ToT
values and saturating at the highest readout value (given by 2$^n$ - 1 where n is the
number of bits). The original raw ToT spectrum is shown in Figure~\ref{fg:raw_TOT} and the spectrum
of a transformed data set, with a lower dynamic range and smaller readout counter
(four bit) is shown in Figure~\ref{fg:raw_TOT_scaled4bit}. The spectrum is truncated at the highest count and all
higher charges are assigned to the highest count. Non-linear calibration effects and
eta corrections were not considered, and are expected not to significantly alter the results.

\begin{figure}[htb]
\centering
\includegraphics[width=0.3\linewidth]{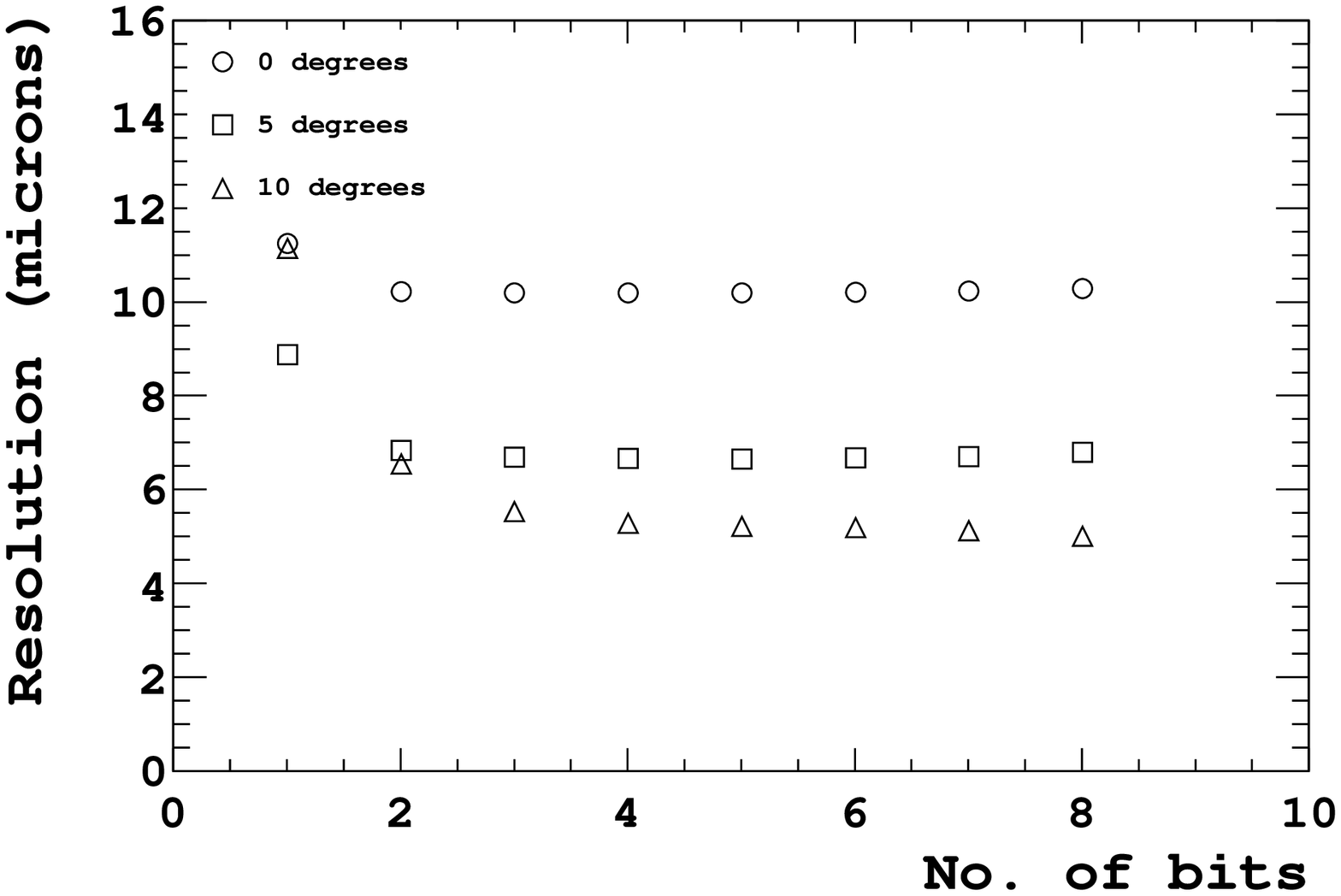}
\includegraphics[width=0.3\linewidth]{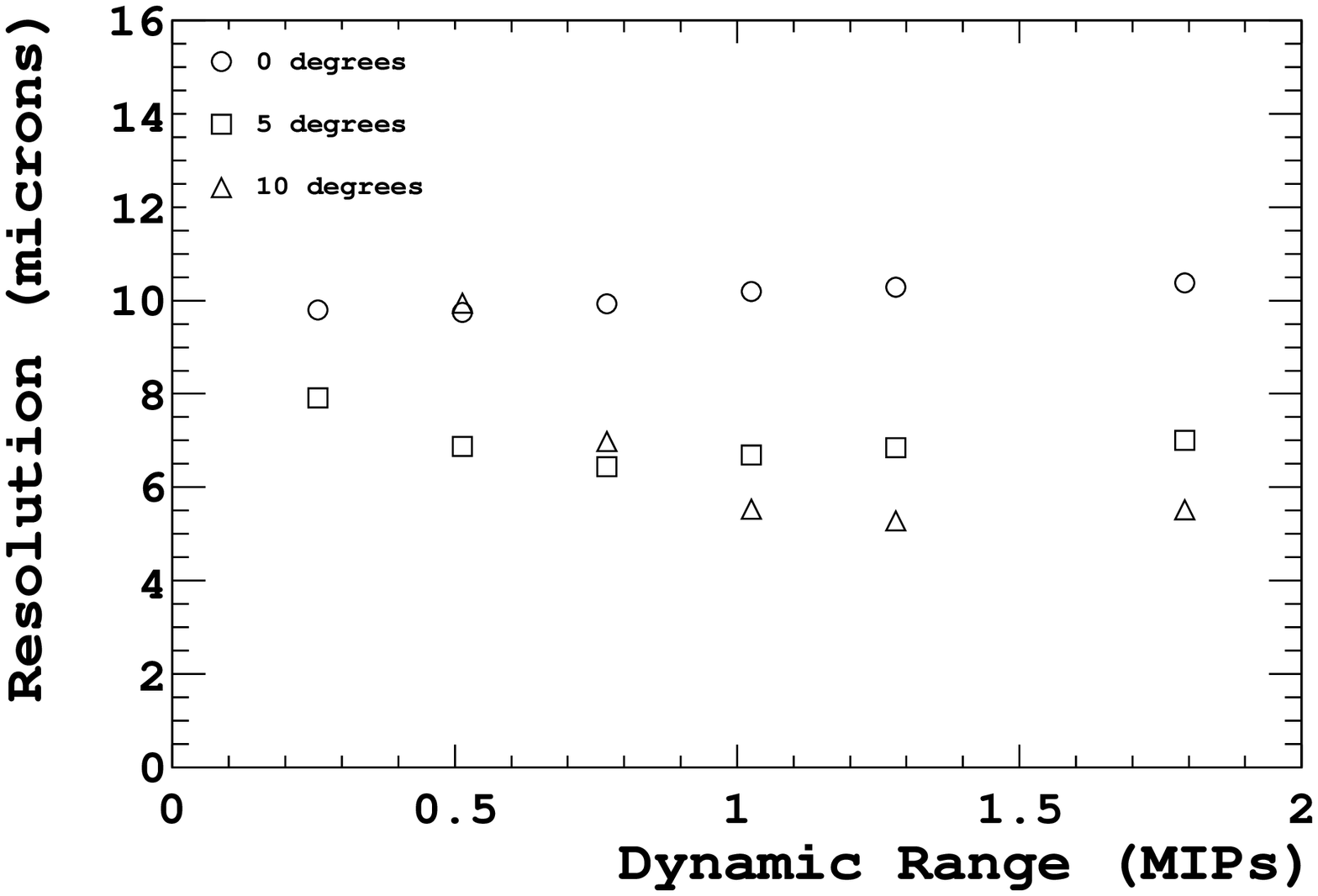}
\includegraphics[width=0.3\linewidth]{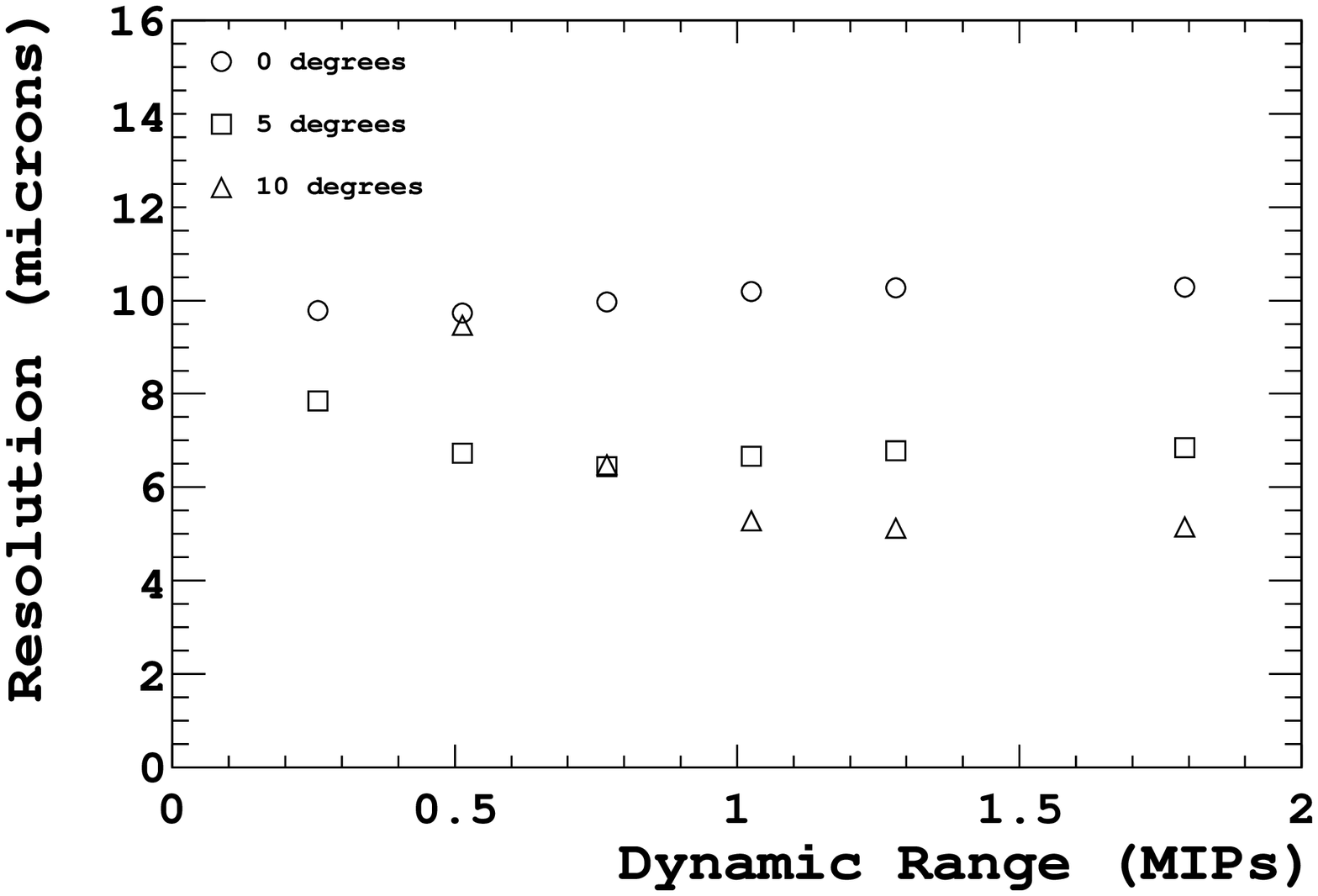}
\caption{From left to right: resolution as a function of
the number of bits read out, resolution versus gain relative to I$_{\textrm{krum}}$=5 for three bit readout,
and resolution versus gain relative to I$_{\textrm{krum}}$=5 for four bit readout.}
\label{fig:bitplots}
\end{figure}

Two main effects were investigated. In the first, the dynamic range was kept constant ($\sim$22.5~ke$^-$)
and the number of bits in the readout counter was varied, effectively changing the charge resolution.
The effect on spatial resolution is shown in Figure~\ref{fig:bitplots} for different angles.
Since the position of one pixel
clusters does not depend on their charge, there is no expected effect from the number of bits. For
multi-pixel clusters however, it is clear that at least a three bit counter is required to maintain good
spatial resolution.

In the second set of plots the dynamic range is varied while keeping the number of readout
bits constant. The resulting spatial resolution is shown for three and four bit readout in Figure~\ref{fig:bitplots}, again at
several angles. It can be seen that a range of at least 1 MIP must be covered, with lower ranges significantly degrading the results.
In summary, the charge conversion cover a range of least 1 MIP with three bit resolution to maintain the best spatial resolution.
\twoepspicx{raw_TOT}{Raw ToT spectrum.}{raw_TOT_scaled4bit}{Scaled ToT spectrum.}
\section{Conclusions}
\label{s:conclusions}

This paper describes the characterisation of Timepix as a charged
particle tracking device.  A particle telescope has been constructed
using $1.4 \times 1.4$ ~cm square sensitive planes consisting of
Timepix and Medipix2 ASICs bonded to planar silicon sensors.
The pixel pitch was $55$~$\mu$m
square.  This telescope is shown to achieve
a track pointing precision of $2.3 \pm 0.1$~$\mu$m and was
operated at a track reconstruction rate of 200~Hz.  Two different
devices were tested at the centre of the telescope.  A Timepix bonded
to a planar sensor was exhaustively tested, and a procedure was
developed to calibrate the sensors with testpulses.  This calibration is particularly
important for charge deposits resulting from the trajectories of
minimum ionising charged particles, which fall at the lower edge of
the sensitivity of the Timepix front-end.  Using the calibrations a precise comparison has
been made between the landau distributions obtained and the expected
values from literature.  The test device was shown to have an
efficiency of greater than $99.5 \%$, and a charge calibration in agreement
with expectations.  The resolution was
extracted as a function of track incident angle, in both the normal and
diagonal directions.  A best resolution of  $4.0 \pm 0.1$~$\mu$m
was measured at the optimum angle of $\sim 10^{\circ}$.  For tracks at
perpendicular incidence the resolution was found to be a strong
function of bias voltage, reaching  $4.4 \pm 0.1$~$\mu$m at 10~V,
the depletion voltage of the device under test, and $10 \pm 0.2$~$\mu$m
at a depletion voltage of 100~V, corresponding to typical
voltages at which the device would have to be operated in LHC
environments.  The performance was also investigated as a function of
the number of bits used in the analogue readout, and the charge
sharing properties described.   The time-stamping performance of the
Timepix was measured and compared to expectations.  A double-sided 3D sensor bonded to a Timepix was also
characterised and the telescope precision used to make detailed
efficiency maps of the pixel cell.  For angled tracks the 3D sensor
achieved equivalent efficiency to the planar sensor.  The resolution
was also extracted and the sensor was seen to produce less charge
sharing than the planar device.  

These studies demonstrate
comprehensively that Timepix is a very suitable device for charged
particle tracking, and its adaptation for use for the LHCb upgrade will be actively
pursued.  The prototype telescope will be upgraded to boost the data acquisition
rate and time stamping peformance and the performance will be
described in a subsequent paper.
\section{Acknowledgments}
The authors would like to thank
Lukas Tlustos,
Winnie Wong,
Timo Tick,
Raphael Ballabriga, and
Sami V\"{a}h\"{a}nen for many useful discussions and their 
invaluable contributions to this paper.  We are very grateful to 
Ian McGill, of the CERN Wire Bonding and Reliability Testing lab, for very
useful discussions and the wire bonding of the Timepix and Medipix assemblies.
The authors are particularly grateful for the support of Stansilav
Pospisil and his group at the IEAP, CTU, Prague, for providing the USB
readouts and Pixelman software.  We would like to thank the operation
team of the CERN SPS for excellent support and delivery of the pion beam.
This work was partially supported and financed by the ICTS(Integrated
Nano-Microelectronics Clean Room)access within the GICSERV Programme
and by the Spanish project FPA2009-13896-C02-02.
One of the authors (R.Plackett) gratefully acknowledges support from
the ACEOLE Marie-Curie FP7 Fellowship scheme.

\end{document}